\documentstyle[10pt,aaspptwo,epsf]{article}

\lefthead{W. L. Holzapfel \ea }
\righthead{$v_r$ A2163 and A1689}

\def\b{{\beta}}
\def\D{{\Delta}}

\def\T{{\Theta}}
\def\ea{{\it et al.} \,}
\def\sz{Sunyaev \& Zel'dovich\,}

\def\s-z{S-Z}

\def\comp{Comptonization}
\def\pr{^{\prime}}
\def\2pr{^{\prime\prime}}
\def\ah{^{\rm h}}
\def\am{^{\rm m}}
\def\as{^{\rm s}}
\def\greatsim{\mathrel{\raise.3ex\hbox{$>$\kern-.75em\lower1ex\hbox{$\sim$}}}}
\def\lesssim{\mathrel{\raise.3ex\hbox{$<$\kern-.75em\lower1ex\hbox{$\sim$}}}}

\begin{document}

\title{Limits on the Peculiar Velocities of Two Distant Clusters Using
the Kinematic Sunyaev-Zel'dovich Effect}

\author{
W.L.~Holzapfel\altaffilmark{1},
P.A.R.~Ade\altaffilmark{2},
S.E.~Church\altaffilmark{3},\\
P.D.~Mauskopf\altaffilmark{3,4},
Y.~Rephaeli\altaffilmark{5},
T.M.~Wilbanks\altaffilmark{6}
and A.E.~Lange\altaffilmark{3}}


\altaffiltext{1}{Enrico Fermi Institute, University of Chicago, Chicago
IL 60637}
\altaffiltext{2}{Department of Physics, Queen Mary and Westfield
College, Mile End Road, London, E1 4NS, U.K.}
\altaffiltext{3}{Department of Physics, Math, and Astronomy, California
Institute of Technology, Pasadena CA 91125}
\altaffiltext{4}{Department of Physics, University of California,
Berkeley, CA 94720}
\altaffiltext{5}{Center for Particle Astrophysics, University of California,
Berkeley, CA 94720}
\altaffiltext{6}{Aradigm Corporation, 26219 Eden Landing Road, Hayward CA
94545}

\begin{abstract}
We report millimeter-wavelength observations of the 
Sunyaev-Zel'dovich (S-Z) effect in two distant galaxy clusters.
A relativistically correct analysis of the S-Z data is combined 
with the results of X-ray observations to determine the radial 
peculiar velocities ($v_r$) of the clusters.
We observed Abell 2163 ($z=.201$) in three mm-wavelength bands
centered at $2.1\,$, $1.4\,$, and $1.1\,$mm.
We report a significant detection of the thermal component of the
S-Z effect seen as both a decrement in the brightness of the CMB at 
$2.1\,$mm, and as an increment at $1.1\,$mm.
Including uncertainties due to the calibration of the instrument, 
distribution and temperature of the IC gas, and astrophysical
confusion, a simultaneous fit to the data in all three bands gives 
$v_r=+490^{+1370}_{-880}\,{\rm kms}^{-1}$ at $68\%$ confidence.
We observed Abell 1689 ($z=.181$) in the $2.1$ and
$1.4\,$mm bands.
Including the same detailed accounting of uncertainty, 
a simultaneous fit to the data in both bands gives
$v_r=+170^{+815}_{-630}\,{\rm kms}^{-1}$.
The limits on the peculiar velocities of A2163 and A1689 correspond 
to deviations from the uniform Hubble flow of $\lesssim 2-3\%$.
\end{abstract}



\section{Introduction}
\label{cintro}
Inhomogeneities in the large-scale mass distribution
of the universe produce, through their gravitational
interaction, a corresponding velocity field.
Measurements of this velocity field can place useful
constraints on cosmological models, including the mass density
of the universe (e.g., \cite{Dekel}).
Galaxy clusters have been shown to be effective tracers of the
large-scale velocity field of the universe
(\cite{Bahcall}; \cite{Gramann}).
Standard methods for determining the radial peculiar velocities ($v_r$)
of clusters difference the velocity determined from the redshift ($z$) with 
that expected for the uniform Hubble flow,
\begin{equation}
v_r=\frac{c}{q_0^2} \left[q_0 z +(q_0 -1)\left(\sqrt{2 q_0 z + 1} -1\right)\right]- H_0 d_L,
\end{equation}
where $H_0$ is the Hubble constant and $q_0$ is the deceleration 
parameter.
The luminosity distance to the cluster ($d_L$) is typically determined with an 
empirical relationship relating the brightness of a standard candle 
to its distance. 
In general, the uncertainties in the distance indicators, and therefore, 
the cluster peculiar velocities, increase linearly with distance.

\begin{figure}[htb]
\plotone{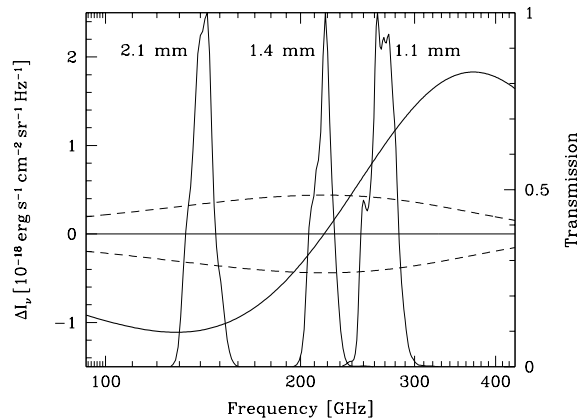}
\caption[]
{Change in CMB Intensity due to the two components of the S-
Z effect. The
non-relativistic approximation to the thermal
component (solid line) is plotted for $y=10^{-4}$ and the
kinematic component (dashed line) is plotted for
($v_r=\pm1000\,{\rm kms}^{-1}$, $\tau=0.01$).
The average transmissions of the three SuZIE spectral
bands (normalized to one) are superimposed.}
\label{szspec}
\end{figure}

The interaction of the Cosmic Microwave Background (CMB) with hot 
intracluster (IC) gas bound
to clusters of galaxies provides a method to directly
determine the radial peculiar velocities of galaxy clusters.
Compton scattering of the CMB 
by hot IC gas -- the Sunyaev-Zel'dovich (S-Z) effect --
gives rise to an observable distortion of the CMB spectrum.
For a general review of the S-Z effect, see 
Rephaeli (1995a).
The change in CMB intensity caused by the scattering
has a thermal component due to the random motions of the 
scattering electrons
(\cite{SZ72}), and a kinematic component due to the bulk
peculiar velocity of the cluster gas (\cite{SZ80}).
The S-Z thermal component produces a decrement in CMB intensity
at low frequency and an increment at high frequency, while
the kinematic component appears as an increment or decrement at all
frequencies.
In Figure~\ref{szspec}, we show the change in the CMB intensity for
the two components of the S-Z effect as a function of frequency.
Accurate measurements  at two or more millimeter/submillimeter wavelengths
can, in principle, separate the two components of the S-Z effect. 

When combined with X-ray measurements, the amplitude of the thermal
component can be used to determine $H_0$
(\cite{Cav79}; \cite{BHA}).
The ratio of the kinematic and thermal component intensities can be
combined with the IC gas temperature to determine the
the radial component of the cluster peculiar velocity
(\cite{SZ80}; \cite{R+L}).
Because the surface brightness of the S-Z effect is independent of
redshift and the derived peculiar velocity is weakly dependent on IC gas 
distribution and temperature,
accurate peculiar velocities can be determined for 
distant clusters which, in general, have poor X-ray data. 

We have developed an instrument and observing strategy which achieves 
high sensitivity to the S-Z effect in three mm-wavelength bands
while minimizing systematic errors (\cite{Holzapfel97a}).
This paper describes the observations and analysis used to place 
upper limits on the peculiar velocities of two distant galaxy clusters.
In Section~\ref{theory}, we discuss the spectra of the S-Z effect
and the determination of peculiar velocities.
The instrument used to make the S-Z observations and its calibration
are described in Section~\ref{mmobs}.
Analysis of the S-Z data is described in Section~\ref{mmanal}.
The results of the observations of A2163 and A1689
are described in Sections~\ref{A2163} and \ref{A1689}.
Astrophysical confusion of the S-Z effect, including the 
contribution of
primary CMB anisotropies, is discussed in Section~\ref{conf}.
We summarize our results and conclusions in Section~\ref{conc}.

\section{Theory}
\label{theory}
\subsection{Relativistic Comptonization in Clusters}
\label{rcomp}
The \sz\ (1972) treatment of CMB Comptonization,
which is commonly used in the analysis of the S-Z effect in clusters,
is valid only at low gas temperatures.
The determination of accurate cluster peculiar velocities using the 
S-Z effect requires the use of the relativistically
correct treatment of the \comp.
We express the relativistically correct intensity 
difference as
\begin{equation}
\D I_{T} =  I_{0} {\int \Psi (x,\, T_e)\, d\tau} ,
\label{ISZT}
\end{equation}
where $I_0=2(kT_0)^3/(hc)^2$, $T_0$ is the CMB temperature,
$d \tau = n_e \sigma_T dl$, and the integral is along the
line of sight through the cluster.
The spectrum is given by
\begin{equation}
\Psi(x,\, T_e)=\frac{x^3}{\left(e^x-1\right)}\left[\Phi\left(x,T_e\right)-1\right] \,,
\end{equation}
where $\Phi(x\,,T_e)$ is a three dimensional
integral fully specified in Rephaeli (1995b).
To facilitate comparison of our results with low frequency
observations, we determine
an effective \comp\ in terms of the measured surface brightness,
\begin{equation}
y = \frac{ \D I_T}{I_0}\, \left[\frac{\int{ \frac{kT_e}{m_ec^2}\, d\tau}}
{\int{ \Psi (x,\, T_e)\, d\tau}}\right]\;.
\label{ycal}
\end{equation}

As we will demonstrate in Section~\ref{scal}, neglecting relativistic 
corrections would lead to considerable errors
in the cluster peculiar velocity.
All results are calculated using
the relativistically correct treatment of the \comp\ unless 
otherwise stated.

\subsection{Peculiar Velocity from S-Z and X-Ray Measurements}
If the IC gas has a bulk velocity (${\vec v_p}$) with
respect to the CMB frame, there is an additional kinematic component
to the S-Z effect,
\begin{equation}
\D I_{K} = -I_{0}\, h(x)\, \int{ n_e \sigma_T \frac{\vec v_p}{c} \cdot \vec{dl}}\, ,
\label{ISZK}
\end{equation}
where the spectrum is given by $h(x)= x^4 e^x/(e^x-1)^2$, identical to
that for a change in the CMB temperature.

Following Birkinshaw, Hughes \& Arnaud (1991) (BHA),
we express the IC gas temperature and density
in terms of the product of a reference value
and a dimensionless form factor (BHA equations 3.3 - 3.4):
\begin{equation}
n_e = n_{e0}f_n(\theta,\,\phi,\,\xi),
\end{equation}
\begin{equation}
T_e = T_{e0}f_T(\theta,\,\phi,\,\xi).
\end{equation}
The angle $\theta$ is measured from a reference line of sight, $\phi$ is the
azimuthal angle about that line of sight, and $\xi$ is the angular
distance along the line of sight.
We extend this treatment to include a term describing the temperature
dependence of relativistic Comptonization,
\begin{equation}
\Psi(x,\,T_e) = \Psi_0(x,\, T_{e0}) f_{\Psi}[f_T(\theta,\,\phi,\,\xi)].
\end{equation}
The expressions for the intensity of the S-Z components,
equations~\ref{ISZT} and \ref{ISZK}, then become:

\begin{equation}
\D I_T \left(\theta,\phi\right) = I_0 n_{e0}
\sigma_T d_A \Psi_0(x,\,T_{e0}) \int{d \xi f_n f_{\Psi}}
\label{SZ-surf}
\end{equation}
\begin{equation}
\D I_K \left(\theta,\phi \right) = -I_{0} n_{e0} \sigma_T  d_A h(x)
\frac{v_r}{c} \int{d \xi f_n} \, .
\label{teqn}
\end{equation}
The structural information for the cluster is contained in
the angular form factors:
\begin{equation}
\T_T(\theta,\, \phi) = \int{d \xi  f_n f_{\Psi}},
\end{equation}
\begin{equation}
\T_K(\theta,\, \phi) = \int{d \xi  f_n }\,.
\end{equation}
The peculiar velocity can be expressed in terms of the
intensity ratio of the two components of the S-Z effect,
\begin{equation}
\frac{v_r}{c}= - \frac{\D I_K \Psi_0(x,\,T_{e0}) \T_T(x,\,T_{e0})}{\D I_T h(x) \T_K}\,.
\end{equation}

In general, measurements are made over finite bands in frequency
which contain contributions from both components of the S-Z effect.
The intensities of the two components of the S-Z effect can be expressed as 
linear combinations of the intensities in the measured bands.
In the case where the IC gas is isothermal, the angular scale factors 
$\Theta_T$ and $\Theta_K$ are identical and results for $v_r$ depend 
only on the ratios of the measured band intensities.

\section{Observations of the S-Z Effect}
\label{mmobs}
\subsection{Instrument}
The S-Z observations were made using the Sunyaev-Zel'dovich
Infrared Experiment (SuZIE) bolometer array at the Caltech
Submillimeter Observatory (CSO) on Mauna Kea.
A detailed description of the SuZIE instrument has been presented
elsewhere (\cite{Holzapfel97a}).
SuZIE is a $2 \times 3$ array of $300\,$mK bolometric detectors optimized
for observations of the S-Z effect in distant $(z > .1)$  clusters of galaxies.
The array consists of two rows separated by $2.2^{\prime}$ in declination; 
each row consists of three co-linear array elements separated by 
$2.3^{\prime}$ in right ascension (RA).
Each array element produces a beam on the sky which is approximately
$1.7\pr$ FWHM.

The spectral responses of the array elements are determined by a
common set of metal-mesh filters.
The filters can be changed for observations in three mm-wavelength
passbands.
The $2.1$ and $1.1\,$mm filter passbands are designed to maximize the
ratio of the S-Z thermal component signal to the sum of the atmosphere
and detector noise.
The $1.4\,$mm filter band is designed to have high sensitivity
to the S-Z kinematic component while constraining the net thermal
component contribution to be near zero.
Due to relativistic corrections, this cannot be done for all IC gas
temperatures; in general, the $1.4\,$mm band contains some
residual thermal component signal (see Section~\ref{scal}).
The average transmission of each of the three SuZIE pass-bands are 
shown in Figure~\ref{szspec}.

Array elements within each of the two rows are electronically
differenced by placing pairs of detectors in AC biased bridge circuits.
The output of each bridge is synchronously demodulated to produce a
stable DC voltage proportional to the instantaneous brightness
difference between the two array elements.
The bolometer differences strongly reject signals common to both elements,
such as fluctuations in the temperature of the $300\,$mK heat sink
and atmospheric emission.
Each row provides three differences, two with beam throws
of $2.3\pr$ and one with $4.6\pr$.

\subsection{Scan Strategy}
\label{scan}
For observations of the S-Z effect, the SuZIE array is used in
drift scanning mode.
The telescope first tracks a position leading the source by a right
ascension offset (RAO).
To begin the scan, the telescope stops tracking and is fixed in place;
the rotation of the earth drifts the source across the array of 
detectors.
The scans are roughly $30^{\prime}$ in length and last 
$\sim 120/({\rm cos} \delta)\,$sec for low declination ($\delta$) clusters.
Between scans, the array is rotated about the optical
axis so that the two rows of the array are kept parallel to the
direction of the scan.
Keeping the telescope fixed while taking data eliminates signals 
due to modulation of the telescope's side-lobes and microphonic 
response of the detectors.
After a scan, the telescope tracks a new position leading the 
source by a different RAO and a new scan is begun.
This cycle, which alternates the RAO between scans is repeated  
indefinitely.
In Section~\ref{baseline}, we use the alternation of the RAO between scans 
to test for an instrumental baseline.

\subsection{Calibration}
\label{cal}
\subsubsection{Flux Calibration}
Observations of planets are used to map the beam shapes of
the instrument and calibrate the responsivity of the
detector elements.
In April 1993 and 1994, Uranus was used to calibrate the
$2.1\,$mm observations.
We used scans of Uranus to calibrate the $1.4\,$mm band 
in May 1994.
In May 1993, observations of Mars were used to
calibrate $1.1\,$mm band observations.
Rotation of the array about the optical axis results in small 
reproducible changes in the instrument's beam-shapes.
For each band, we mapped the beam shapes over the range of 
dewar rotation angles for which we observed the source; the 
calibrations computed for the range of rotation angles at which
we observed the sources change by less than $5\%$.

The brightness of Uranus is found from a third order polynomial
fit to the measured brightness temperature as a function of
wavelength (\cite{Griffin}).
The brightness temperature of Mars is taken from 
Orton \ea (1986).
The authors assign $\pm 6\%$ uncertainty to the brightness of Uranus,
most of which arises from the $\pm 5\%$ uncertainty in the
absolute brightness of Mars from which Uranus is calibrated.
Combining the uncertainties in the
measured beam-shapes and absolute brightness of the planetary
calibrator, the uncertainty in the calibration of the
instrument to surface brightness in each of the spectral
bands, is estimated to be $\sim \pm 8\%$.

We use these results to calculate the contribution of the uncertainty 
in flux calibration
to the uncertainty in the peculiar velocity determined from measurements 
in two spectral bands.
Because the peculiar velocity can be expressed as a ratio of 
measured brightnesses, a calibration error which is common across the bands, 
such as the uncertainty in the absolute calibration of the planetary 
calibrator is likely to contribute, makes no contribution to the uncertainty. 
We have calculated the flux calibration error assuming the worst 
(and unlikely)
case that the $<5\%$ uncertainty in the measured beam-shapes is 
anticorrelated between bands.
For the combination of the $2.1\,$ and $1.1\,$mm data, we determine
the contribution to the uncertainty in the peculiar 
velocity to be $\D v_r \lesssim \pm 140(kT_e/10\,{\rm keV})\, {\rm kms}^{-1}$.
For the combination of the $2.1\,$ and $1.4\,$mm data, the
contribution to the uncertainty is 
$\D v_r \lesssim \pm 6(kT_e/10\,{\rm keV})\,{\rm kms}^{-1}$.
The comparatively large uncertainty determined for the $2.1$ and $1.1\,$mm data is 
because, to determine $v_r$, we are (in effect) differencing 
two large thermal component signals, each with a proportional
flux calibration uncertainty.
For the combination of the $2.1\,$ and $1.4\,$mm data, the uncertainty is
roughly proportional to the small signal in the $1.4\,$mm band.
For a peculiar velocity determination  using the $2.1$, $1.4$, and $1.1\,$mm 
bands, the uncertainty in the flux calibration contributes an 
uncertainty $\D v_r \sim \pm (6-140)\left(T_e/10\,{\rm keV}\right)\, 
{\rm kms}^{-1}$, depending on the relative significance of the data 
in the three bands.
In our accounting of the uncertainties we assume the worst case, 
and use the upper limit.
In this analysis, we have assumed that $5\%$ of $v_r$ 
is a negligible contribution to the total uncertainty, true for 
the results of this work. 

\subsubsection{Spectral Calibration}
\label{scal}
The spectral responses of the array elements,
including detailed checks for out of band leaks,
are measured using a Fourier transform spectrometer.
Because the beams of the array look through different portions
of the filter at different angles, 
there are slight differences between their spectral bands.
Small deviations of the $1.4\,$mm band from the null frequency
of the thermal effect produce large offsets in 
the derived peculiar velocities.
To demonstrate the size of this effect, we calculate the peculiar
velocity that one would determine by assuming all the signal in the
$1.4\,$mm band is due to the kinematic effect,
\begin{equation}
\frac{\D v_r}{c} = \frac{\int d\nu \, \Psi(\nu,\, T_e)\,f(\nu)}{\int d\nu \, h(\nu)\, f(\nu)}\,.
\end{equation}
In the non-relatistic approximation, the thermal effect signal
in the average $1.4\,$mm band is nearly zero.
In Table~\ref{dspec}, the scatter of the non-relativistic results about 
zero are due to the array element differences.
The results are listed for $T_e=10\,$keV;
the offsets scale linearly with the assumed $T_e$.
These significant differences in the spectral response of the
array elements are taken into account by individually calibrating each 
of them to the two components of the S-Z effect.

\begin{table*}
\begin{center}
\begin{tabular}{ccccc}
\multicolumn{5}{c}{Thermal Effect Signal in $1.4\,$mm Band $[ {\rm
kms}^{-1}]$}\\\tableline\tableline
  &  \multicolumn{3}{c}{ $T_e\,[{\rm keV}]$} & \\
Detector & $8.2$ & $12.4$ & $13.3$ & non-relativistic ($10\,$ keV) \\\tableline
$1$ & $229$ & $517$ & $581$ & $-25.8$\\
$2$ & $399$ & $764$ & $837$ & $210.1$\\
$3$ & $299$ & $616$ & $687$ & $72.7$\\
$4$ & $113$ & $351$ & $406$ & $-186$.0\\
$5$ & $356$ & $702$ & $773$ & $151.5$\\
$6$ & $161$ & $420$ & $478$ & $-119.5$\\\tableline
average & $261$ & $550$ & $627$  & $19.6$ \\
\end{tabular}
\end{center}
\caption[]{Residual thermal effect signal in the $1.4\,$mm band
calibrated in terms of an ``offset'' in peculiar velocity.
Results are shown for the three principal IC gas temperatures
discussed in this paper. We also show the results using the
non-relativistic S-Z spectra for $10\,$keV IC gas.}
\label{dspec}
\end{table*}

Relativistic corrections to the S-Z spectrum lead to significant thermal 
component signal in the $1.4\,$mm band.
In Table~\ref{dspec}, we show the error in peculiar velocity that would be 
incurred by
ignoring relativistic corrections to the S-Z effect for the three 
principal gas temperatures considered in this work. 
In the non-relativistic approximation, the average $1.4\,$mm band is 
centered at the null of the S-Z thermal effect.
Relativistic corrections shift the null to higher frequencies; the change 
in frequency is linearly dependent on the IC gas temperature (\cite{Rephaeli95a}).
The S-Z spectra, as sampled by the narrow $1.4\,$mm band 
(See Figure \ref{szspec}), increases nearly linearly with frequency.
The amplitude of the thermal component also increases nearly linearly
with increasing IC gas temperature.
Combining these effects, we find that neglecting relativistic corrections 
leads to an error in the peculiar velocity determined form the average
$1.4\,$mm band of $\D v_r\approx +360(kT_e/10{\rm keV})^2$.
This is corrected for by calibrating with the correct relativistic 
spectra.

There is uncertainty in the results of the laboratory
spectroscopy.
From repeated spectroscopy runs, we conservatively estimate
the uncertainty in the array element central frequencies 
to be $< \pm 1\%$.
When the peculiar velocity is determined from the 
combination of the $2.1$ and $1.1\,$mm band data, the uncertainty
in $v_r$ is $\pm 137(kT_e/10{\rm keV})\,{\rm kms}^{-1}$. 
The steeper slope of the thermal spectra near the null 
leads to a slightly larger uncertainty of
$\pm 196(kT_e/10{\rm keV})\,{\rm kms}^{-1}$
when the velocity is determined from 
combination of the $2.1$ and $1.4\,$mm bands.
All three bands are used to determine the peculiar
velocity of A2163; we conservatively adopt the larger
of the uncertainties determined from two bands.

\section{S-Z Data Analysis}
\label{mmanal}
This section describes the reduction and analysis of the S-Z scans.
The analysis of the A2163 $2.1\,$mm data set is presented in detail in 
H97b.
The initial reduction of all the S-Z data is identical to 
that treatment.
We outline the analysis method, with particular emphasis on the determination
of accurate confidence intervals from multiparameter fits to 
the S-Z scans.

\subsection{Data Set}
\label{mmset}
Each of the two rows of the SuZIE array consists of three detectors; 
$s_1$, $s_2$ and $s_3$.
Detector signals are differenced in pairs (in hardware) to form three
difference signals; $d_{12}=s_1-s_2$, $d_{23}=s_2-s_3$, and
$d_{31}=s_3-s_1$.
These differences correspond to angular chops of $2.3^{\prime}$,
$2.3^{\prime}$, and $4.6^{\prime}$ respectively.
The two $2.3^{\prime}$ differences are differenced
to form a triple beam chop (TBC), for example, 
\begin{equation}
t_{123} = d_{12}-d_{23}= s_1 - 2s_2 +s_3 \,.
\end{equation}
The data set consists of $4$ difference signals: $d_{31}$, $t_{123}$,
$d_{64}$, and $t_{456}$ corresponding to the $4.6^{\prime}$ difference
and TBC for each of the two rows of detectors.
We refer to these four difference signals as $d_{k}$, 
where $k$ ranges from 1 to 4. 
We also ompute the average signal of the undifferenced detectors,
\begin{equation}
s=\frac{1}{6}\sum_k^6 s_{k}\,.
\end{equation}
The average single detector signal is used as a monitor of the absolute
atmospheric emission and to remove any residual common mode response
from the detector differences.

We clean the raw data of transients due to the
interaction of cosmic rays with the detectors.
Less than $5\%$ of the data are identified as contaminated by cosmic rays.
The raw data are binned into $3\,$s bins corresponding to $15$
$5\,$Hz samples or $0.75^{\prime}\,{\rm cos}(\delta)$ on the sky.
Samples flagged as bad due to cosmic rays are left out when the
bin averages are computed; bins with more than half of
their samples flagged are not used in the analysis.
For each scan j, and bin i, the binned differential and 
average single detector signals 
are expressed as $d_{kji}$ and $s_{ji}$.

\subsection{S-Z Surface Brightness Model}
\label{mod}
The S-Z surface brightness profile of A2163 has a FWHM $\approx 5\pr$,
which is comparable to the largest beam-throw of SuZIE.
In order to accurately determine the peak surface brightness,
we must simulate the observation of the extended source.
Models for the surface brightness of the S-Z thermal and kinematic
components are constructed from the X-ray surface brightness determined
density profile and the assumption of some thermal structure.
We express the surface brightness morphology in terms of dimensionless
form factors normalized to one at peak brightness:
\begin{equation}
S_{Ti}(\theta, \phi)= \frac{\D I_{T\nu_{i}}(\theta,\,\phi)}{\D I_{T\nu_{i}}(0,0)} =
\frac {\int f_n(\theta,\phi,\xi)\, f_{\Psi i}(\theta,\phi,\xi) \,d \xi} {\int
f_n(0,0,\xi)\, f_{\Psi i}(0,0,\xi)\, d\xi}\,,
\end{equation}
\begin{equation}
S_{Ki}(\theta, \phi)= \frac{\D I_{K\nu_{i}}(\theta,\,\phi)}{\D I_{K
\nu_{i}}(0,0)} =
\frac {\int f_n(\theta,\phi,\xi)\, d\xi} {\int
f_n(0,0,\xi)\, d\xi}\,.
\end{equation}
To take differences in the array elements into account, $S_{K,Ti}$
are evaluated at the S-Z intensity weighted band center for each 
array element i.
In the case of an isothermal IC gas,
$S_{Ti}(\theta,\,\phi) =S_{Ki}(\theta,\,\phi)$ for all i.

The form factor for the desired S-Z component ($S_{K,Ti}$) is convolved
with beam-maps constructed from the voltage response of 
the detectors to scans over planets, $V_{Pi}$.
This forms a template for the response of each detector to
the assumed surface brightness of the S-Z components,
\begin{eqnarray}
\lefteqn{sm_{K,Ti}(\theta)=}\\
& & \hspace{-10 pt} \frac{\D I_{K,Ti}}{I_{Pi}} \frac{1}{\Omega_P} \int V_{Pi}
(\theta-\theta^{\prime},\,\phi^{\prime}) \,
S_{K,T}(\theta^{\prime},\,\phi^{\prime})\, d\theta^{\prime} d\phi^{\prime}\,, \nonumber
\label{ssb}
\end{eqnarray}
where $\Omega_P$ is the solid angle subtended by the planetary
calibrator.
This expression assumes the planetary calibrator to be much smaller
than the beams on the sky, true for Mars and Uranus.  
To calibrate the model, we determine the ratio of
the relativistically correct S-Z brightness to the
brightness of the planet,
\begin{equation}
\frac{I_{K,Ti}}{I_{Pi}} = \frac{\int f_i(\nu) I_{K,T\nu}\, d\nu}
{\int f_i(\nu) I_{P \nu}\, d\nu} \,,
\end{equation}
where $f_i(\nu)$ is the measured spectral response and
$I_{P\nu}$ is the intensity of the planetary calibrator.
The S-Z brightness is calculated for $y=1$ and $v_r=1\,{\rm kms}^{-1}$,
so the results of the model fits are in convenient units. 

The models for the response of the single array elements are 
differenced to create
models for the response of the detector differences.
The $\sim .05^{\prime}$ resolution differential source models are 
binned identically
to the scan data to determine the model signal for each of
the $\sim .75^{\prime}$ data bins.
The binned differential models are designated as $m_{ki}(RA)$, where $k$ is
one of the four detector differences, $i$ is the position (by
bin number) in the scan, and RA is a offset of the model from the
nominal X-ray determined position.

\subsection{Coadded Data}
\label{Coadd}
We coadd the difference signals to create high sensitivity 
scans of the differential surface brightness as a function of RA.
For each difference channel k, and scan j, we clean the data by 
removing the best fit linear baseline and residual common-mode 
signal ($\alpha_{kj}s_{ji}$),
\begin{equation}
x_{kji} = d_{kji} -\alpha_{kj}s_{ji} -a_{kj} -i\,b_{kj}\,.
\label{clean}
\end{equation}
The residual common-mode signal is the small signal in the detector differences
that is correlated with the average single detector signal.
Removing this signal improves the rejection of the differential signals
to common mode atmospheric noise. 
In H97b, we show that removing it has no systematic effect on the fit 
results.

The value of each coadded bin $x_{ki}$, is given by the weighted sum 
of the value of this bin in each of $N_s$ scans,
\begin{equation}
x_{ki}= \frac{\sum\limits_{j=1}^{N_s} 
\frac{x_{kji}}{RMS_{kj}^2}}{\sum\limits_{j=1}^{N_s}\frac{1}{RMS_{kj}^2}}\,.
\label{coadd}
\end{equation}
Each bin is weighted by the residual RMS of the scan,
\begin{equation}
RMS_{kj}^2 = \frac{\sum\limits_{i=1}^{N_{b}}x^2_{kji}}
{\left( N_{b} -1 \right)}\,,
\label{cowt}
\end{equation}
where $N_b$ is the number of bins in a scan.
The uncertainty in the value of each bin is determined from the weighted
dispersion 
of the value of that bin about the mean determined from $N_s$ scans,
\begin{equation}
\sigma_{ki}= \sqrt{\frac{\sum\limits^{N_s}_{j=1}{\frac{(x_{ki}
-x_{kji})^2}{RMS_{kj}^2}}}{(N_{s}-1)\sum\limits_{j=1}^{N_s} 
\frac{1}{RMS_{kj}^2}}}\,.
\label{coerr}
\end{equation}

The best fit peak \comp\ ($y_0$) and isothermal model position ($RA$) are
found by minimizing the
$\chi^2$ of the fit to all four difference signals for each of the
coadded $2.1\,$mm data sets,
\begin{equation}
\chi^2 =
\sum_{k=1}^{4}\sum_{i=1}^{N_b}\frac{\left(x_{ki}-y_0\,m_{Tki}(RA)-a_{k}-i\,b_{k}\right)^2}{\sigma_{ki}^2}\,.
\label{cafit}
\end{equation}
The linear baseline is unconstrained to make sure that
our removal of a linear baseline in equation~\ref{clean} did not effect 
the signal.
For each value of RA, we re-bin the $.05^{\prime}$ resolution source model
to determine the model signal for each of the $\sim .75^{\prime}$
data bins.

If the source contributes significantly in a single scan, then a scan in
which the noise and source conspire to produce a low RMS and corresponding
low signal will be weighted higher.
Coadding the scans using equations~\ref{clean}--\ref{cowt}
could introduce a bias in the amplitude of the coadded scans.
In H97b, we eliminate this bias by recomputing the coadded bins 
with each bin weighted on the residual RMS
with the average source model removed.
We do this for all the coadded data presented in this paper as well, although 
it is unlikely to be necessary 
for the extremely low signal to noise $1.4$ and $1.1\,$mm scans.

\subsection{Single scan fits}
\label{Scanfits}
The determination of accurate uncertainties for fits to the coadded 
scans is difficult due to the presence of atmospheric noise which
is correlated between bins and detector differences.
Ignoring this correlation leads to a significant overestimate of
the significance of the fit results.
In this section, we use the distribution of the single scan fit amplitudes  
to determine accurate uncertainties for model amplitudes.
These results are then used to correct the error in the coadded fit 
uncertainties due to correlated sky noise.

The noise in scans taken at different times is uncorrelated.
Each scan can be treated as an independent measurement
of the source amplitude. 
We determine the mean single-scan peak \comp\ for a given
observation by averaging the values $y_{0j}$ for each of the $N_s$
scans,
\begin{equation}
y_0 = \frac{\sum\limits_{j=1}^{N_s}{\frac{y_{0j}}{{\sigma_{yj}}^2}}}{\sum\limits_{j=
1}^{N_s}{\frac{1}{{\sigma_{yj}}^2}}}\,,
\label{ymean}
\end{equation}
where $\sigma_{yj}$ is the change in $y_{0j}$ corresponding to 
$\D \chi^2=1$ in the model fit.
The weighted dispersion of the scan amplitudes about the
mean is used to estimate the uncertainty in the determination of
the mean,
\begin{equation}
\sigma_y = \sqrt{
\frac{\sum\limits_{j=1}^{N_s}{\frac{(y_0-y_{0j})^2}{{\sigma_{yj}}^2}}}{(N_s-1)\sum\limits_{j=1}^{N_s}{\frac{1}{{\sigma_{yj}}^2}}}}\,.
\label{yerror}
\end{equation}

In H97b, we plot the distribution of single scan
fit amplitudes for all of the $2.1\,$mm scans across
A2163.
As in that example, the distribution of scan fit amplitudes 
for all the data sets considered here are well approximated by a Gaussian. 
The width of the distribution is a function of the amount of sky
noise and is correlated with the RMS of the scans.
Using the weighting scheme described in this section, we narrow 
the distribution slightly by decreasing the weight of the scans with large
sky noise.

\subsection{Confidence intervals}
\label{like}
From the single-scan fits we have an accurate measurement of
the uncertainty in the source amplitude 
for each data set, $\sigma_y$.
We compare this to the uncertainty ($\sigma_{yca}$) determined from
the fits to the coadded data (equation~\ref{cafit}; $\D \chi^2 =1$).
We compute a scaling factor to correct for the
underestimation of the bin uncertainties for each
data set, m.
\begin{equation}
\gamma_m = \frac{\sigma y}{\sigma_{yca}}\,.
\end{equation}
Depending on the details of the atmosphere at the time of
the measurement, $\gamma_m = 1.4 - 2.0$.
Each coadded scan is individually scaled
so that the uncertainties in the coadded scan fit amplitudes
are equal to the single-scan fit uncertainties.
Fits to the coadded data (for similar model 
shapes and positions) will then result in accurate uncertainties
for the fit parameters. 
Essentially, we have reduced the number of degrees of freedom to
account for the correlated noise.
The resulting confidence intervals are a factor of $1.4$ to
$2.0$ larger than those found from the uncorrected data.
All confidence intervals in this paper are computed using coadded data that
has been corrected for correlated sky noise in this way.

We determine confidence intervals for the source position, 
peculiar velocity, peak \comp\, and isothermal $\beta$ model parameters
using a maximum likelihood estimator, 
\begin{equation}
L_{m}(\zeta) = \prod_k^4\prod_{i=1}^N \frac{1}{(2\pi\, \gamma_m \,
\sigma_{ki})^{1/2}}
\, {\rm exp} \left[\frac{
-\zeta_{ki}^2(\,y,\,z)}{2\,(\gamma_m\,\sigma_{ki})^2}\right]\,,
\end{equation}
where $\zeta_k = x_{ki}-y_0\,m_{Tki}(RA,\,\theta_c,\,\beta)\\-v_r\,m_{Kki}(RA,\,\theta,\,\beta)-a_k -i\,b_k$ is the coadded scan with the kinematic and thermal
component source models removed.
The coadded scans corresponding to different RAOs and 
frequency bands are treated as independent;
the likelihood for simultaneous fits of 
$k$ scans for each of $m$ uncorrelated data sets is given by, 
\begin{equation}
L(\zeta) = \prod_m \prod_k L_{mk}(\zeta)\,.
\end{equation}

To make the problem computationally tractable we determine
the likelihood for only two interesting parameters at a time. 
The likelihoods are determined for a necessarily large grid in the 
parameter space with resolution: $\Delta RA=.05^{\prime}$, 
$\D v_r =10\,{\rm kms}^{-1}$, and $\D y_0=2\times 10^{-6}$.
Assuming no peculiar velocity, fits to the X-ray derived isothermal
thermal component model determine the source position and amplitude.
The resulting position is insensitive to the values of the peculiar 
velocity and $\beta$ model parameters.
For the rest of the analysis, we fix the source position to its best fit value.
The isothermal $\beta$ model parameters are determined with the 
source amplitude free to vary.
We do this to check that the assumed model is consistent
with the S-Z data.
Finally, we fix the position and $\beta$ parameters for the  
the thermal and kinematic component models and determine
the likelihood for $v_r$ and $y_0$. 

The likelihood grids are converted to confidence regions and
intervals in the interesting parameters.
Invoking Bayes' theorem and assuming a uniform prior, the probability
that the two fit parameters $(y,\,z)$ fall within a region R is given by
\begin{equation}
P(R) = \frac{\int \int_R  L(y,\,z) \,dydz}{\int_{-\infty}^{\infty} \int_{-\infty}^{\infty} L(y,\,z)\,dydz}\,.
\end{equation}
The confidence interval corresponding to a probability $P_0$
is given by the region $R_0$ such that
$P(R_0) = P_0$ and $L[(y,\,z) \in R_0] \geq L[(y,\,z)\not\in R_0]$.
We determine the likelihood in a single parameter by
marginalizing the likelihood over the uninteresting parameters,
\begin{equation}
L(z) = \int^{\infty}_{-\infty}  L(y,\,z)\,dy\,.
\end{equation}
The probability that the parameter y falls within the interval
$I = [z_1,\,z_2]$ is given by, 
\begin{equation}
P(I) = \frac{\int_{I} L(z)\,dz}{\int^{\infty}_{-\infty} L(z)\,dz}\,.
\end{equation}
The confidence interval corresponding to a probability $P_0$
is given by $I_0=[z_1,\,z_2]$ such that
$P(I_0) = P_0$ and $L(z \in I_0) \geq L(z \not\in I_0)$.
All the $68\%$ confidence levels quoted in Sections~\ref{A2163} 
and \ref{A1689} are calculated this way unless otherwise stated.

\subsection{Baseline}
\label {baseline}
We have carefully designed our instrument and scan
strategy to eliminate systematic errors.
However, it is possible that the instrument could introduce some
systematic signal similar to the expected cluster signal.
To test for the presence of this type of instrumental baseline,
we gather ``blank sky'' scans across regions of sky free of
known sources.
These data are analyzed exactly as the source data;
the presence of an instrumental baseline would result in the
determination of a non-zero signal for these regions.

In April 1994, we accumulated $\sim 15$ hours of $2.1\,$mm data on two
patches of sky free of known sources.
The analysis of these data are described in 
Holzapfel \ea (1997b).
Fitting these data to the isothermal thermal effect model for A2163
with the position in the scan fixed at the best fit position from the
source observations, we determine $y_0 = 1.55 \pm 2.13 \times 10^{-5}$.
In May 1993, we accumulated $\sim 7.2$ hours of $1.1\,$mm
data on three patches of sky free of known sources.
Fitting these data to the isothermal thermal effect model for A2163,
we determine $y_0 = -.21 \pm 2.17 \times 10^{-4}$.
In May 1994, we accumulated $\sim 16.3$ hours of $1.4\,$mm
data on two patches of sky free of known sources.
Fitting these data to the kinematic component source model for A2163
and assuming $\tau =.015$,
we find $v_r=  +760 \pm 1050 \, {\rm kms}^{-1}$.
Therefore, we detect no significant baseline in any of the
three spectral bands. 

We could subtract these scans from the source scans 
to remove a possible instrumental baseline.
Unfortunately, due to shorter integrations and poorer weather,
the ``blank sky'' data in the $1.4$ and $1.1\,$mm bands is less
sensitive than the source data; the subtraction of these scans 
would significantly increase the uncertainty in the fit results.
Another problem with this procedure is that the baseline and source
data, although gathered at similar azimuth and zenith angle, are
gathered at different times.
Any baseline signal, because it is not significant in a single scan,
must be correlated in time between several scans.
However, this does not guarantee an instrumental baseline will be constant
over the course of the night.

We have devised an observation strategy which allows us to check for,
and remove, any time correlated baseline.
Differencing scans taken adjacent in time at
two different RAOs allows the subtraction of any common baseline
while retaining most of the expected source signal.
Each pair of scans are differenced by subtracting their raw
$5\,$Hz sampled time streams.
The data are then analyzed exactly as described in
Sections~\ref{mmanal}--\ref{like}.

We construct a model for the RAO differenced surface brightness by
subtracting two models with the appropriate RAO difference between them.
We chose the RAO difference ($6^{\pr}$) to be large enough so that
(for most clusters) the
two models have little overlap in time and therefore will not subtract
much signal when differenced.
The RAO differenced model is fit to the 
RAO differenced data to determine the source amplitude and 
position.
In all cases the results agree, within the quoted statistical error,
with those found from fits to both RAOs.
The lack of any significant change when the scans are differenced
indicates that there is no significant instrumental baseline
in the data.
We use the difference between the results of the RA differenced 
and standard analysis as an estimate of
the uncertainty due to the possibility of an instrumental baseline.

\section{A2163}
\label{A2163}
\subsection{X-ray Surface Brightness}
Abell 2163 is a distant $(z=.201)$ X-ray luminous cluster of galaxies
with the distinction of having the highest X-ray temperature of
any cluster yet observed.
A2163 was the target of pointed observations by the GINGA, ROSAT
and ASCA satellites.
A detailed analysis of each of these observations has been published
(\cite{Arnaud92}; \cite{Elbaz}; \cite{Markevitch96}).
The ROSAT/PSPC was used to determine the spatial dependence of the X-ray
surface brightness.
The peak surface brightness was found at
$16\ah\,15\am\,46\as;\;-06^{\circ}\,09\pr\,02\2pr$ (J$2000$).
The radial profile was determined by summing
annuli about this position, with significant emission detected
up to $18\pr$ from the position of the peak.
Assuming the IC gas to be isothermal, the surface brightness profile
was fit with the combination of an isothermal $\beta$
model and a constant background.
The density profiles are assumed to have the form of a spherically
symmetric $\beta$ model,
\begin{equation}
f_n(\theta,\phi,\xi) = \left[1+\left(\frac{\theta^2 + \phi^2 + \xi^2
}{\theta_c^2}\right)\right]^{-\frac{3}{2}\beta}\,.
\label{bmod}
\end{equation}
The best fit model parameters were found to be
$\beta = .616_{-.009}^{+.012}$ and $\theta_c = 1.20 \pm 0.05\pr$ at
$68.3\%$ confidence (\cite{Elbaz}).

\subsection{Thermal Structure}
\label{Hybrid}
In H97b, the thermal structure of the IC gas in A2163 is discussed
in detail, here we summarize only the essential points.
A combined analysis of the ASCA/GIS+SIS and GINGA spectral data gives 
temperatures: $13.3$, $13.3$, and $3.8\,$keV in three radial bins of
$0-3$, $3-6$, and $6-13$ X-ray core radii (\cite{Markevitch96}).
We use a simple two parameter model to describe the 
measured temperature profile of A2163. 
The model consists of an isothermal central region extending to
$\theta_{iso}$ beyond which the
temperature decreases according to a polytropic model with index $\gamma$,
\begin{equation}
\begin{array}{l}
        T_e(\theta) = \left\{
        \begin{array}{ll}
                T_{iso} & \theta \leq \theta_{iso} \\
                T_{iso}\left[\frac{1+\left(\theta/\theta_c\right)^2}{1+\left(\theta_{iso}/\theta_c\right)^2}\right]^{-\frac{3}{2}
                \b \left(\gamma -1 \right)} & \theta > \theta_{iso}\,.
        \end{array}
        \right.
\end{array}
\end{equation}
The central temperature is $T_{e0}\approx 13.3^{+2.8}_{-1.8}\,$keV,
where the error bars enclose the $68\%$ confidence intervals for 
both the GINGA and ASCA results, each analyzed independently under 
the assumption of the ASCA thermal structure (H97b).
We adopt thermal model parameters 
$\theta_{iso}=4.0\,\theta_c$ and $\gamma = 2.0$
and reanalyze the X-ray data to
determine consistent density model parameters $\beta=.64$ 
and $\theta_{iso}=1.26$.
These results are used to create a model for the thermal component 
surface brightness that includes the thermal structure.

In the case when the IC gas is assumed to be isothermal,
we adopt a temperature which is the emission-weighted
average temperature 
of the above results.
For an isothermal IC gas, ${\bar T_e} \approx 12.4^{+2.7}_{-1.7}\,$keV,
where the error bars enclose the $68\%$ confidence intervals for 
both the GINGA and ASCA results, each independently analyzed assuming 
an isothermal IC gas.
The results in this section are for fits to the isothermal
source model unless otherwise stated.

\subsection{$2.1\,$mm Observations}
\label{2163.21}
We have observed A2163 at $2.1\,$mm in two periods;
April 23-26 1993 for a total
of $16$ hours and April 4, 9, 10, and 11 for $8$ hours.
The analysis of these data to determine the Hubble
constant is presented in
Holzapfel \ea (1997b).

\begin{figure}[tb]
\plotone{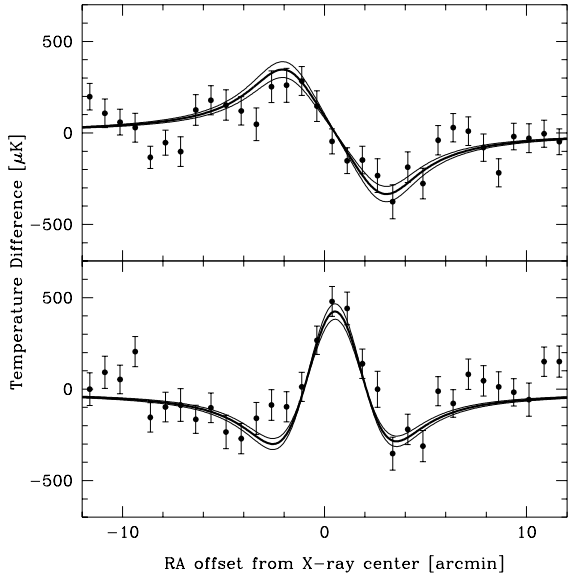}
\caption[]
{Coadded data for all April 1994 scans across A2163 at $2.1\
,$mm.
The coadded scans from the two RAOs have been offset and
added to create a single coadded scan for each difference.
Both the $4.6^{\prime}$ difference (upper panel) and
TBC (lower panel) data are shown for the row
passing directly over the peak X-ray surface brightness.
The heavy line is the best fit isothermal model, and the lighter lines
represent the $1\sigma$ errors.
The scans are calibrated in terms of a Rayleigh-Jeans temperature
difference that completely fills one of the beams.}
\label{sb2163.21}
\end{figure}

\begin{figure}[htb]
\plotone{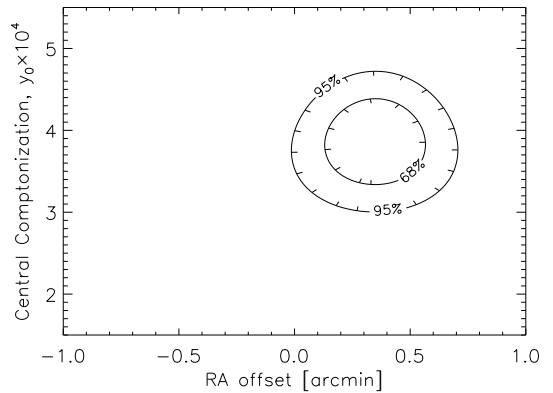}
\caption[]
{Isothermal thermal component model model fit to the coadded
 $2.1\,$mm
data for A2163.
Assuming no peculiar velocity, the contours enclose the
approximate $68.3\%$
and $95.4\%$ confidence regions in peak \comp\ and
RA offset from the X-ray center.}
\label{p2163.21}
\end{figure}

The data corresponding to each observation run and RAO are 
combined to make coadded scans. 
In Figure~\ref{sb2163.21}, 
we plot the April 1994 coadded $4.6\pr$ and
TBC scans over A2163 with the best fit isothermal model overlaid.
Assuming the peculiar velocity to be zero, the coadded data scans are 
simultaneously fit to the corresponding isothermal source 
models with the peak S-Z surface brightness position 
and peak \comp\ free to vary.
In Figure~\ref{p2163.21}, we show the $68.3$ and $95.4\%$ confidence 
regions for the source position and amplitude ($RA$ and $y_0$).
The $68.3\%$ confidence limits for the fit parameters
considered individually are
$\D RA=.35 \pm .14^{\prime}$ and $y_0 = 3.73 \pm 0.35 \times 10^{-4}$.
Including the uncertainties in the ROSAT/PSPC and SuZIE pointing,
the positions of the peak X-ray and S-Z surface brightnesses are
consistent (\cite{Holzapfel97b}).

\begin{figure}[hbtp]
\plotone{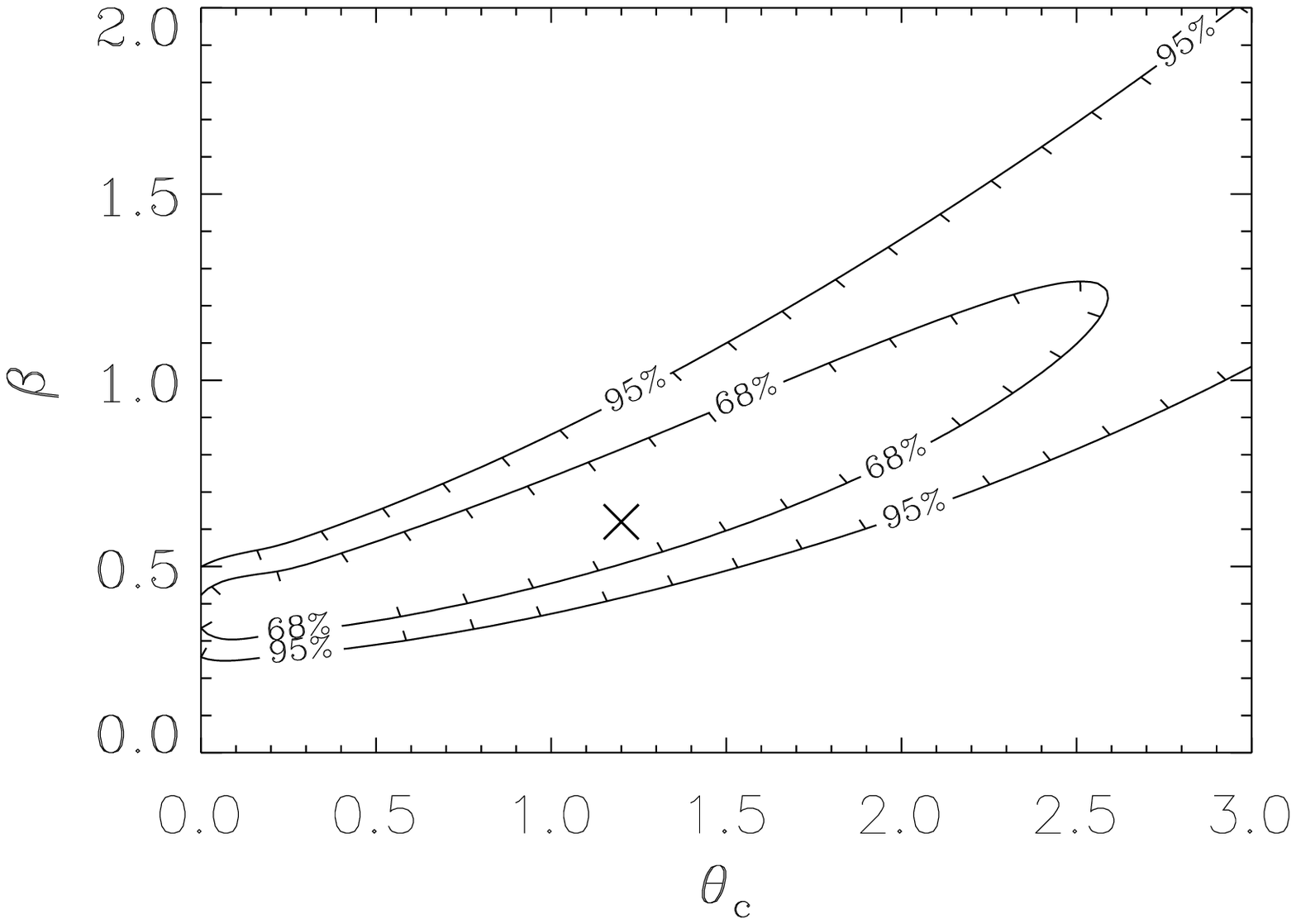}
\caption[]
{Isothermal density model parameter fits to the
coadded data scans across A2163.
Contours for $\theta_c$ and $\beta$ are shown at approximate
ly
$68\%$ and $95\%$ confidence.
The ``X'' marks the best fit isothermal model parameter valu
es
from the X-ray analysis.}
\label{rb2163}
\end{figure}

Fixing the position to the best fit value, we fit isothermal
$\beta$ models for the
surface brightness distribution to the coadded scans.
In Figure~\ref{rb2163},
we plot approximate confidence regions
in $\theta_c$ and $\beta$.
Technically, the data lack sufficient sensitivity to
determine proper confidence regions in position and amplitude;
we approximate the $68.3$ and $95.4\%$ confidence regions
as corresponding to $\D \chi^2 = 2.3$ and $6.17$ respectively.
The results of the S-Z fits are consistent with those of the 
X-ray derived density parameters and an isothermal IC gas. 
In H97b, we found that the S-Z data did not have sufficient 
sensitivity to discriminate between an isothermal IC gas and the 
thermal structure measured with ASCA.

\subsection{$1.1\,$mm Observations}
\label{2163.11}
We observed A2163 on May 1-4 1993 for a total of $19.7$ hours.
Scans were $30^{\prime}$ long with one row passing over
the X-ray center and the other $2.2\pr$ to the south.
The RAO of the X-ray center from the start of the scan was
alternated between $12.5^{\prime}$ and $18.5^{\prime}$
in sequential scans.
The $1.1\,$mm observations are dominated by sky noise
and the quality of the data varies greatly with atmospheric
conditions.

\begin{figure}[htb]
\plotone{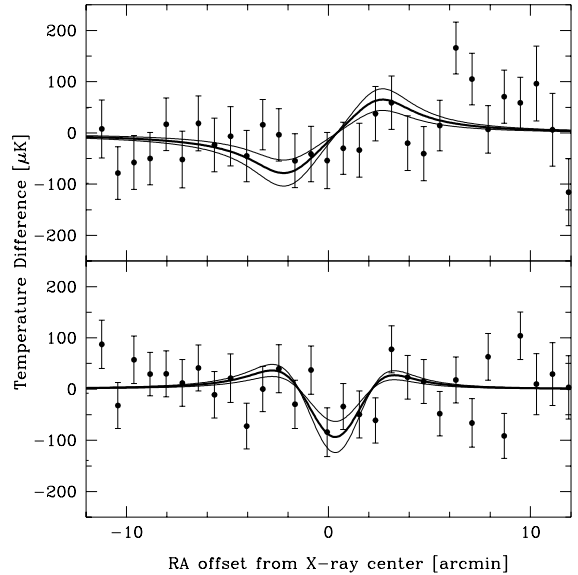}
\caption[]
{Coadded data for all May 1993 scans across A2163 at $1.1\,$mm.
The scans for the two RAOs have been offset and
added to create a single coadded scan for each difference.
Both the $4.6^{\prime}$ difference (upper panel) and
TBC (lower panel) shown for the row
passing directly over the peak X-ray surface brightness.
The heavy line is the best fit isothermal model, and the lighter lines
represent the $1\sigma$ errors.
The scans are calibrated in terms of a Rayleigh-Jeans temperature
difference that completely fills one of the beams.}
\label{sb2163.11}
\end{figure}

\begin{figure}[hbtp]
\plotone{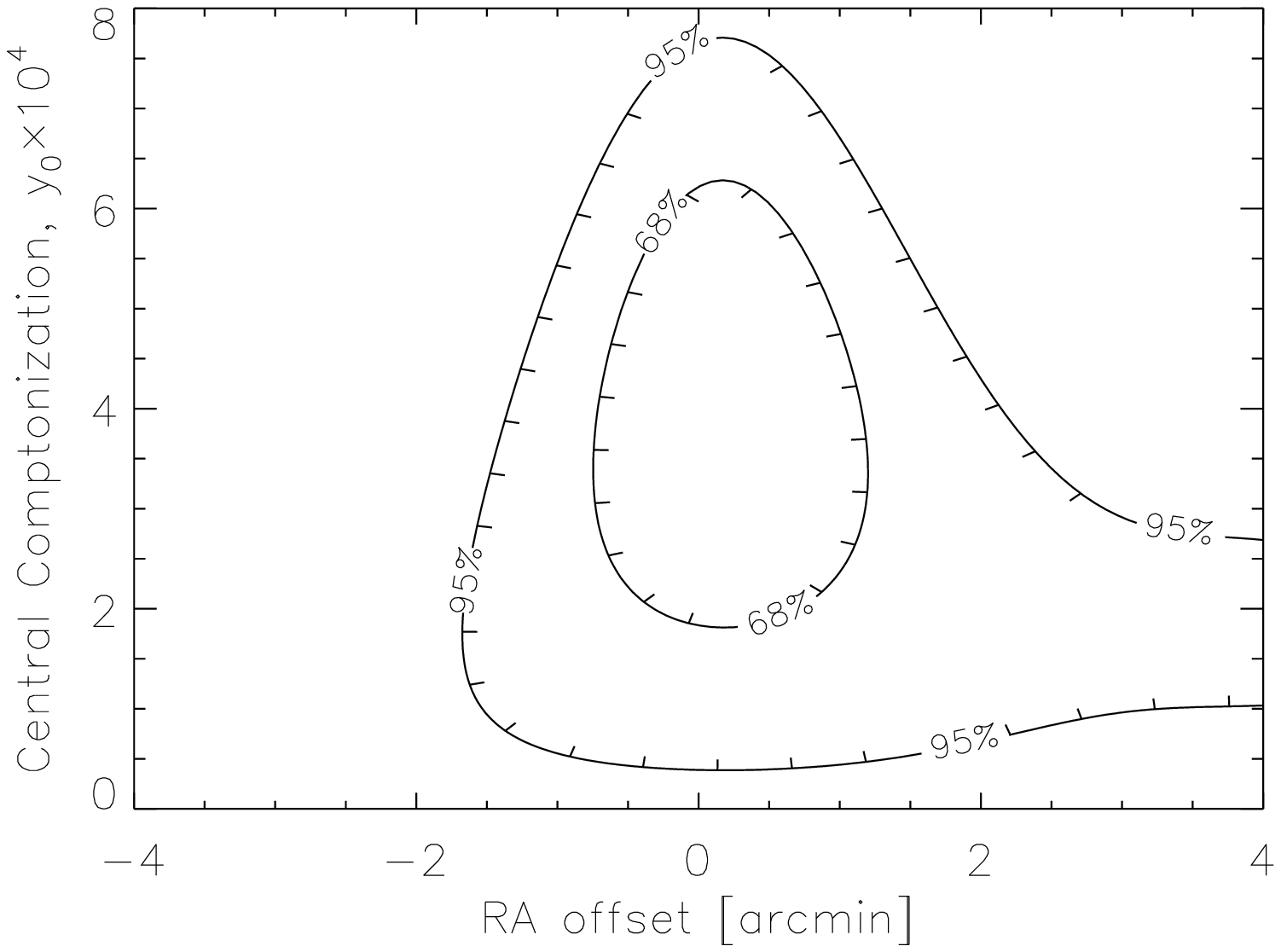}
\caption[]
{Isothermal thermal component model model fit to the coadded $1.1\,$mm
data for A2163.
Assuming no peculiar velocity, the contours enclose the
approximate $68.3\%$
and $95.4\%$ confidence regions in peak \comp\ and
RA offset from the X-ray center
(Note: scale is different than Figure~\ref{p2163.21}).}
\label{p2163.11}
\end{figure}

In Figure~\ref{sb2163.11}, we show the coadded $4.6\pr$ and
TBC scans over A2163 with the best fit isothermal model overlaid.
Assuming the peculiar velocity to be zero, the coadded data scans are 
simultaneously fit to the corresponding isothermal source 
model with the position and peak \comp\ free to vary.
In Figure~\ref{p2163.11}, we show the approximate confidence regions in RA
and $y_0$ computed for the fits of the isothermal source model to the 
$1.1\,$mm coadded data.
The approximate single-parameter
$68.3\%$ $(\D \chi^2 =1.0)$ confidence intervals 
are $\D RA = .17 \pm .59^{\prime}$ and
$y_0 = 4.05 \pm 1.47 \times 10^{-4}$.
Both the amplitude and position of the peak S-Z surface brightness
are consistent with the results of the $2.1\,$mm analysis.

\nopagebreak

\subsection{$1.4\,$mm Observations}
\label{2163.14}
\begin{figure}[hbtp]
\plotone{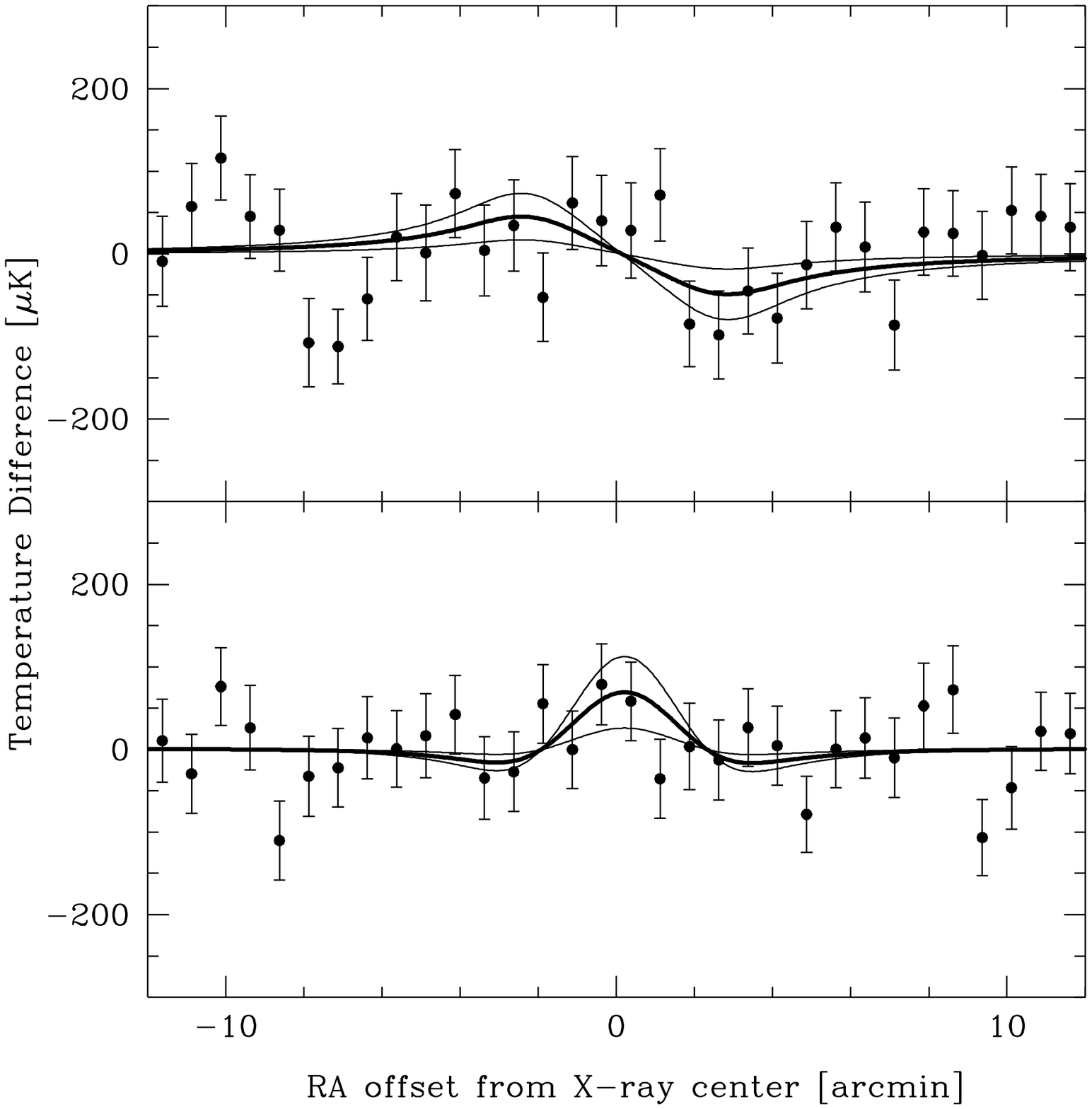}
\caption[]
{Coadded data for all May 1994 scans across A2163 at $1.4\,$mm.
The coaded scans from the two RAOs have been offset and
added to create a single coadded scan for each difference.
Both the $4.6^{\prime}$ difference (upper panel) and
TBC (lower panel) are shown for the row
passing directly over the peak X-ray surface brightness.
The heavy line is the best fit isothermal model, and the lighter lines
represent the $1\sigma$ errors.
The scans are calibrated in terms of a Rayleigh-Jeans temperature
difference that completely fills one of the beams.}
\label{sb2163.14}
\end{figure}

We observed A2163 in the $1.4\,$mm band on the nights of
May 5, 6, 7, and 8 1994 for a total of $\sim 12.3\,$hours.
Scans were $30^{\prime}$ long with one row passing over
the X-ray center and the other $2.2\pr$ to the south.
The RAO of the X-ray center from the start of the scan was
alternated between $12^{\prime}$ and $18^{\prime}$
in sequential scans.
In Figure~\ref{sb2163.14}, we show the coadded $4.6\pr$ and
TBC scans.
Fitting the data with a model for the kinematic component, we
determine a marginally significant signal, the bulk of which
is due to the residual thermal component signal in the
$1.4\,$mm band.

\subsection{Simultaneous Fits}
\label{sim2163}
\subsubsection{2 Band Fits}
We simultaneously fit the 
$2.1\,$ and $1.1\,$mm coadded scans
with the S-Z thermal and kinematic component isothermal models generated
for each of the two bands.
The models are fixed to the best fit position from the $2.1\,$mm
scans.
We determine $v_r=-120^{+1350}_{-1060}\, {\rm kms}^{-1}$ and 
$y_0=3.92 \pm .59 \times 10^{-4}$ at $68.3\%$ confidence.

We also simultaneously fit the 
$2.1\,$ and $1.4\,$mm coadded scans
with the S-Z thermal and kinematic component isothermal models generated
for each of the two bands.
The models are fixed to the best fit position from the $2.1\,$mm
scans. 
We determine $v_r = +990^{+1730}_{-1190}\, {\rm kms}^{-1}$
and $y_0 = 3.45 \pm .63 \times 10^{-4}$ at $68.3\%$ confidence. 

\subsubsection{$2.1\,$mm, $1.4\,$mm and $1.1\,$mm Bands}
\begin{figure}[htb]
\plotone{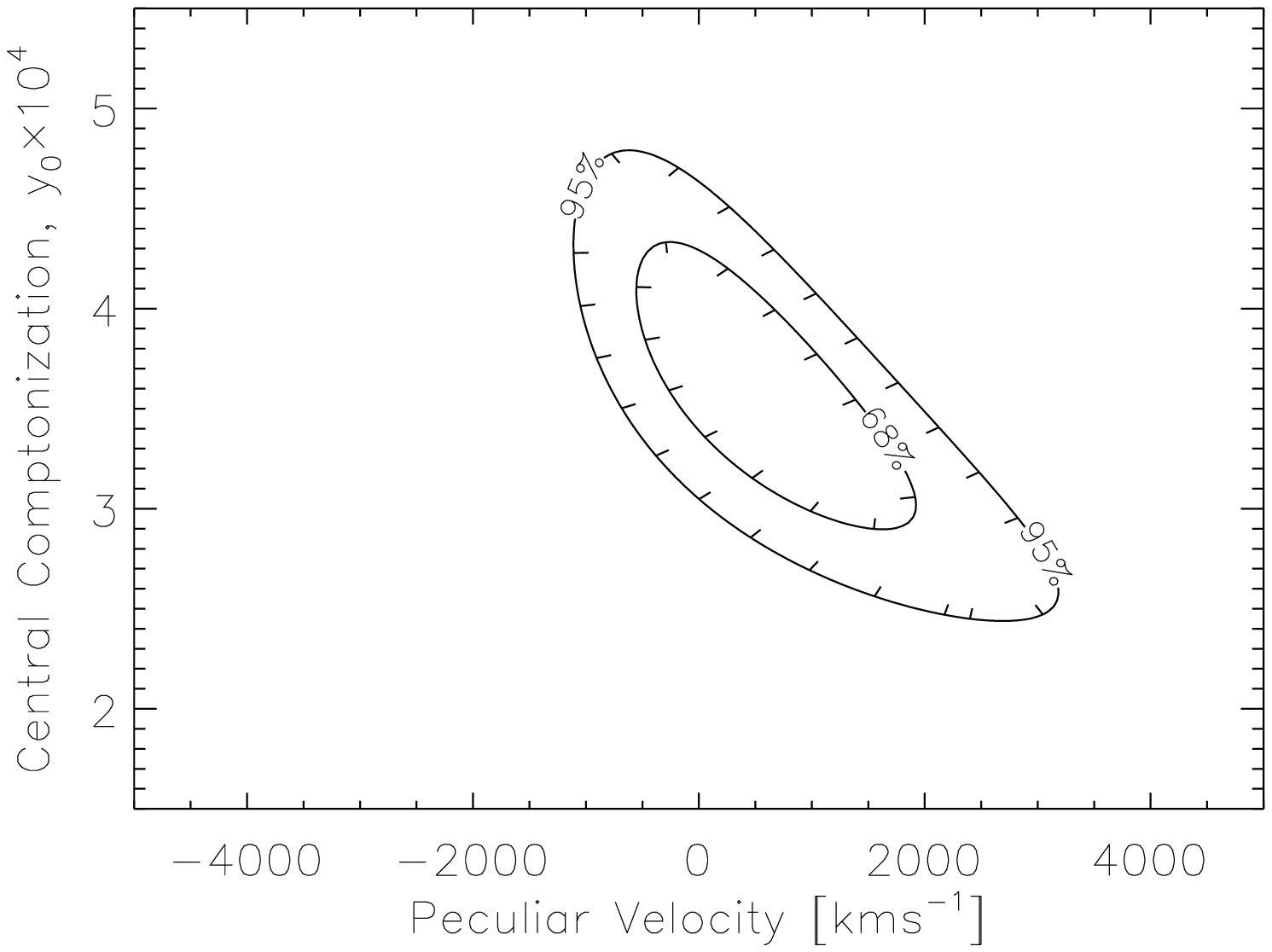}
\caption[]
{Results of fitting the
$2.1$, $1.4$, and $1.1\,$mm coadded data for A2163
to isothermal ($T_e=12.4\,$keV) source models for the two
components of the S-Z effect.
The $68.3\%$ and $95.4\%$ confidence regions in
peak \comp\ and radial component of the peculiar
velocity are shown.}
\label{yv2163}
\end{figure}

We simultaneously fit isothermal ($T_e=12.4\,$keV) models for the 
kinematic and thermal components of the S-Z effect to the
$2.1\,$, $1.4\,$ and $1.1\,$mm coadded scans.
The model positions are fixed to the best fit positions from the 
$2.1\,$mm fits; the model amplitudes, $y_0$ and $v_r$, are 
left as free parameters. 
We plot the $68.3\%$ and $95.4\%$ 
confidence regions for $y_0$ and $v_r$ in Figure~\ref{yv2163}.
The $68.3\%$ confidence intervals for the fit parameters
considered individually are 
$v_r = +490^{+910}_{-730}\, {\rm kms}^{-1}$
and $y_0 = 3.62 \pm .48 \times 10^{-4}$.

We have repeated this analysis for the range of allowed
density models; the results are unchanged within the precision
of the grid in $v_r$ ($10\,{\rm kms}^{-1}$). 
This analysis is repeated using the RAO differenced scans;
we compare the results of the two analysis methods
in Table~\ref{pvbt}.
The sensitivity in $v_r$ is much worse for the RAO differenced data.
This is due to the large extent of the A2163 surface brightness.
A larger RAO difference would reduce the amount of subtracted
signal and increase the sensitivity of the RAO differenced data.

\renewcommand{\arraystretch}{1.25}
\begin{table*}[htb]
\begin{center}
\begin{tabular}{cccccc}
\multicolumn{6}{c}{Peculiar Velocities: Standard and RAO Differenced}\\\tableline\tableline
Cluster & Thermal Structure & $kT_{e0}\,[{\rm keV}]$ & analysis & $y_0\times 10^4$ & $v_r\, [{\rm
 kms}^{-1}]$
\\\tableline
A2163 & Isothermal & $12.4$  & Both RAOs & $3.62\pm.49$ & $+490^{+910}_{-730}$ \\
A2163 & Isothermal & $12.4$  & Differenced & $3.44\pm.72$ & $+1390^{+2350}_{-1440}$ \\
\\
A2163 & ASCA & $13.3$ & Both RAOs & $2.96\pm.39$ & $+560^{+960}_{-800}$ \\
A2163 & ASCA & $13.3$ & Differenced & $2.34\pm.57$ & $+1460^{+2420}_{-1520}$ \\
\\
A1689 & Isothermal & $8.2$ & Both RAOs & $3.43\pm.59$ & $+170^{+760}_{-570}$ \\
A1689 & Isothermal & $8.2$ & Differenced & $3.00\pm.73$ & $+310^{+1050}_{-700}$\\
\\
\end{tabular}
\end{center}
\caption[]{Comparison of peculiar velocities determined with the
standard analysis and fits to the RAO differenced scans.
All results are at $68\%$ confidence.
The differences between the results represent our best limits on
possible contribution due to an instrumental baseline.
The sensitivity is considerably worse for the RAO differenced scans,
especially for A2163, which is more extended than A1689
and has more signal subtracted in the differencing.}
\label{pvbt}
\end{table*}
\renewcommand{\arraystretch}{1.00}

We have repeated the determination of $v_r$ assuming two other 
values for the isothermal IC gas temperature which roughly
span the $68.3\%$ confidence interval for $T_e$.
The peculiar velocities determined assuming $T_e=10\,$ and $15\,$keV 
are listed in Table~\ref{pvtt}.
The effects of the implied change in optical depth and 
realtivistic corrects cancel.
The contribution of the uncertainty in the IC gas 
temperature to the uncertainty in peculiar velocity 
($\D v_r < \pm 20\,{\rm kms}^{-1}$)
is negligible compared to the statistical uncertainty in 
the measurement.

\renewcommand{\arraystretch}{1.25}
\begin{table*}[htb]
\begin{center}
\begin{tabular}{ccccc}
\multicolumn{5}{c}{Peculiar Velocities: IC gas Temperature}\\\tableline\tableline
Cluster & Thermal Structure & $kT_{e0}\,[{\rm keV}]$ & $y_0\times 10^4$ &
$v_r\, [{\rm kms}^{-1}]$ \\\tableline
A2163 & Isothermal & $10.0$ & $3.50\pm.47$ & $+510^{+740}_{-610}$ \\
A2163 & Isothermal & $12.4$ & $3.62\pm.49$ & $+490^{+910}_{-730}$ \\
A2163 & Isothermal & $15.0$ & $3.70\pm.50$ & $+480^{+1030}_{-830}$ \\
A2163 & Hybrid(ASCA) & $13.3$ & $2.96\pm.39$ & $+560^{+960}_{-800}$ \\
\\
A1689 & Isothermal & $5.0$ & $3.31\pm.52$ & $+190^{+470}_{-350}$ \\
A1689 & Isothermal & $8.2$ & $3.43\pm.59$ & $+170^{+760}_{-570}$ \\
A1689 & Isothermal & $10.0$ & $3.51\pm.55$ & $+130^{+900}_{-670}$ \\
\\
\end{tabular}
\end{center}
\caption[]{Peculiar velocities assuming several values of the
IC gas temperature for each cluster.
For A2163, we also show the results for the Hybrid(ASCA) thermal
structure.
Results are for fits to the coadded data at $68\%$ confidence.}
\label{pvtt}
\end{table*}
\renewcommand{\arraystretch}{1.00}

\subsection{$2.1$, $1.4$, and $1.1\,$mm Fits With Thermal Structure}
\label{simtherm}
ASCA observations of A2163 suggest that the IC gas is not isothermal. 
We have repeated the simultaneous fits to the A2163 S-Z data for the
case in which the cluster is assumed to have the ASCA determined
thermal structure (Section~\ref{Hybrid}).
Using the hybrid IC gas temperature model and the consistent 
density model, we create models for the surface brightness of the 
thermal and kinematic components of the S-Z effect.
For a single free parameter, we find
$v_r = +560^{+960}_{-800}\, {\rm kms}^{-1}$
and $y_0 = 3.00 \pm .40 \times 10^{-4}$
at $68.3\%$ confidence.
These values are compared with the isothermal results 
in Table~\ref{pvtt}.
We have repeated this analysis with the RAO differenced scans;
We compare the results of the two analysis methods
in Table~\ref{pvbt} 
Although there is considerable uncertainty in the thermal
structure of the cluster, the uncertainty it contributes
to the determination of the peculiar velocity is negligible 
in comparison to the statistical uncertainty.

\section{A1689}
\label{A1689}
\subsection{X-ray Results}
Abell 1689 is a distant $(z=.181)$ and X-ray luminous cluster
of galaxies.
It has been observed with the ROSAT, GINGA, and ASCA
satellites.
ROSAT/PSPC observations determined the peak of the X-ray surface
brightness to be centered at
$13\ah\,11\am\,29.1\as;$ $-01^{\circ}\,20\pr\,41\2pr$ (J$2000$) with
significant emission detected to $8.5^{\prime}$ (\cite{Daines}).
The X-ray surface brightness contours are nearly circular with 
ellipticity $< 0.1$.
Assuming the density of the IC gas to be described by an
isothermal beta model (equation~\ref{bmod}),
Daines \ea (1995) determine 
$\theta_c = 1.13 \pm 0.12^{\prime}$ and
$\beta = 0.78 \pm 0.03$ at $90\%$ confidence.
From the observed radial velocities of the cluster member galaxies,
it appears there are two to three substructures superimposed along the
line of sight.
While this may complicate the use of this cluster to
determine the Hubble constant, it has no effect on the determination
of the peculiar velocity.

The presence of a cooling flow
($\dot{M}\approx 500\, {\rm M}_{\sun}{\rm yr}^{-1}$)
within the central core of the cluster
complicates the X-ray spectral analysis.
Daines \ea (1995) use the combined analysis of
GINGA and ROSAT spectra to determine $T_e \sim 9\,$keV outside
the cooling flow.
Analysis of ASCA spectra has also been used to determine the
temperature of the IC gas. 
Assuming the gas to be isothermal outside
of the central core region the temperature is determined
to be $T_e = 8.2\pm1.0\,$keV (\cite{Bautz}).

\subsection{$2.1\,$mm Observations}
We observed A1689 in the $2.1\,$mm band on the nights of
April 6, 7, and 9 1994 for a total of $\sim 8.5\,$hours.
Scans were $30^{\prime}$ long with one row passing over
the X-ray center and the other $2.2\pr$ to the south.
The RAO of the X-ray center from the start of the scan was
alternated between $12^{\prime}$ and $18^{\prime}$
in sequential scans.

\begin{figure}[htb]
\plotone{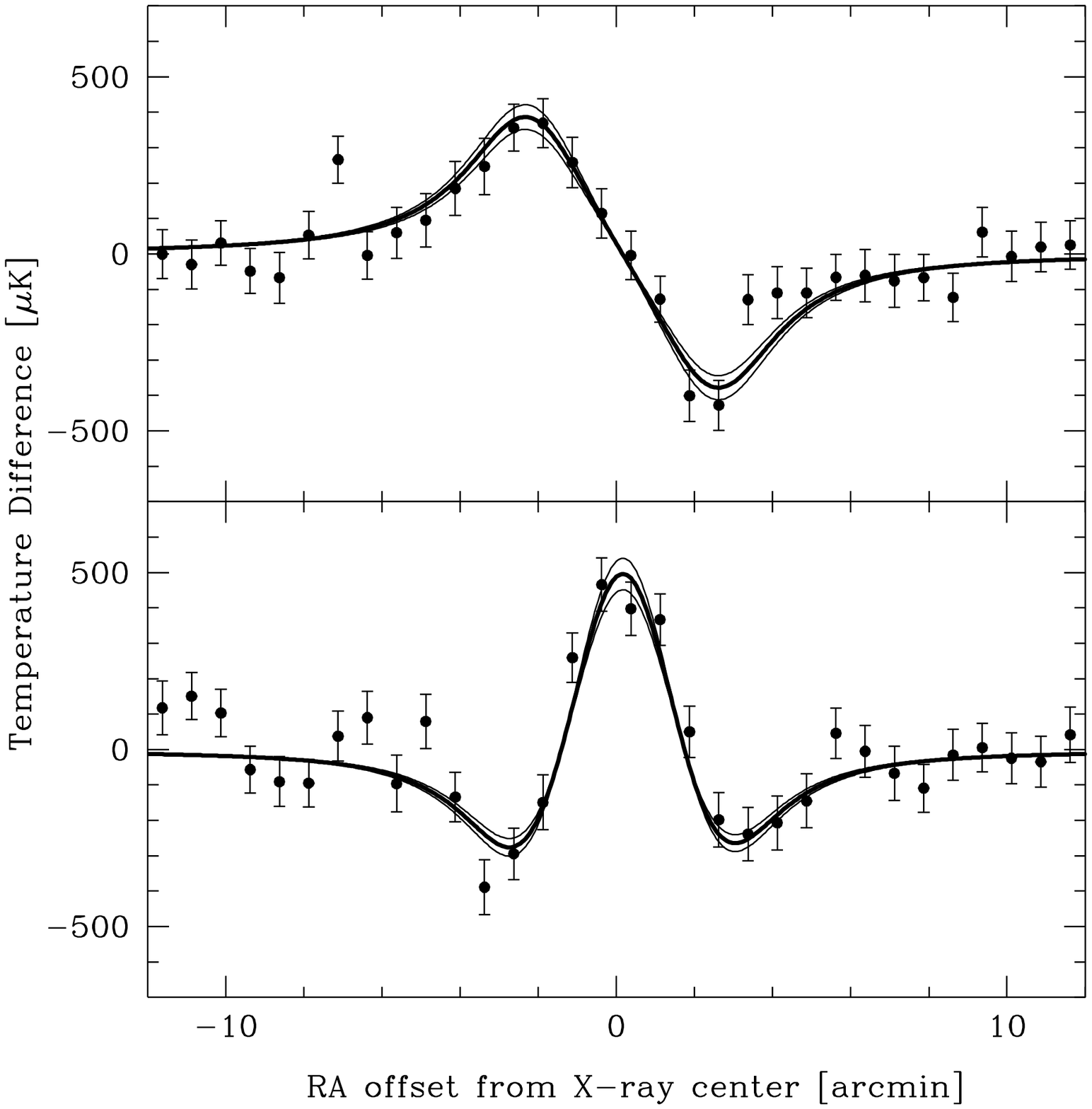}
\caption[]
{Coadded data for all April 1994 scans across A1689 at $2.1\,$mm.
The coaded scans from the two RAOs have been offset and
added to create a single coadded scan for each difference.
Both the $4.6^{\prime}$ difference (upper panel) and TBC(lower panel)
data are shown for the row
passing directly over the peak X-ray surface brightness.
The heavy line is the best fit isothermal model, and the lighter lines
represent the $1\sigma$ errors.
The scans are calibrated in terms of a Rayleigh-Jeans temperature
difference that completely fills one of the beams.}
\label{sb1689.21}
\end{figure}

\begin{figure}[htb]
\plotone{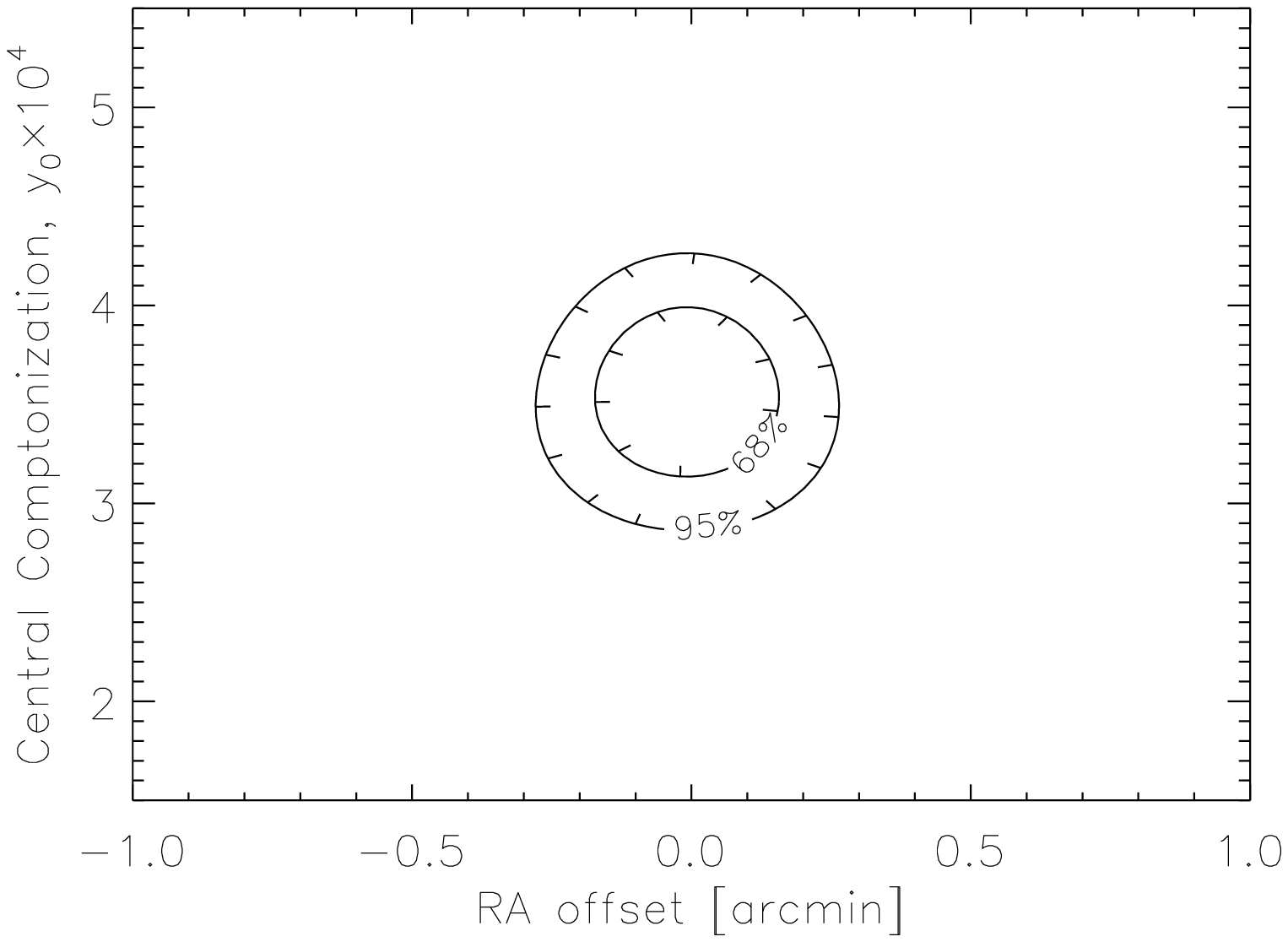}
\caption[]
{Isothermal thermal component model fit to the
coadded $2.1\,$mm data for A1689.
Assuming no peculiar velocity, the contours enclose the $68.3\%$
and $95.4\%$ confidence regions in peak \comp\ and
RA offset from the X-ray center.}
\label{p1689.21}
\end{figure}

In Figure~\ref{sb1689.21}, we show the coadded $4.6\pr$ and
TBC scans over A1689 in Figure~\ref{sb1689.21}.
Assuming no peculiar velocity, we fit
the isothermal model for the S-Z thermal component to
the coadded scans.
In Figure~\ref{p1689.21},
we plot the $68.3$ and $95.4\%$
confidence regions for $y_0$ and RA.
The $68.3\%$ confidence intervals for the position and peak \comp\
are $\Delta {\rm RA} = -.01\pm .11^{\prime}$ and 
$y_0=3.55\pm .29 \times 10^{-4}$.
The positions of the X-ray and S-Z surface brightnesses 
are coincident.

Combining the peak \comp\ with the X-ray determined 
IC temperature, we determine the optical depth, $\tau \sim 0.022$.
This extremely high value of optical depth is exceptional
and suggests the extension of this cluster along the line of sight.
The sensitivity of our measurements to cluster peculiar velocity
scale linearly with $\tau$ making this an ideal cluster for the
determination of the peculiar velocity.

\begin{figure}[htb]
\plotone{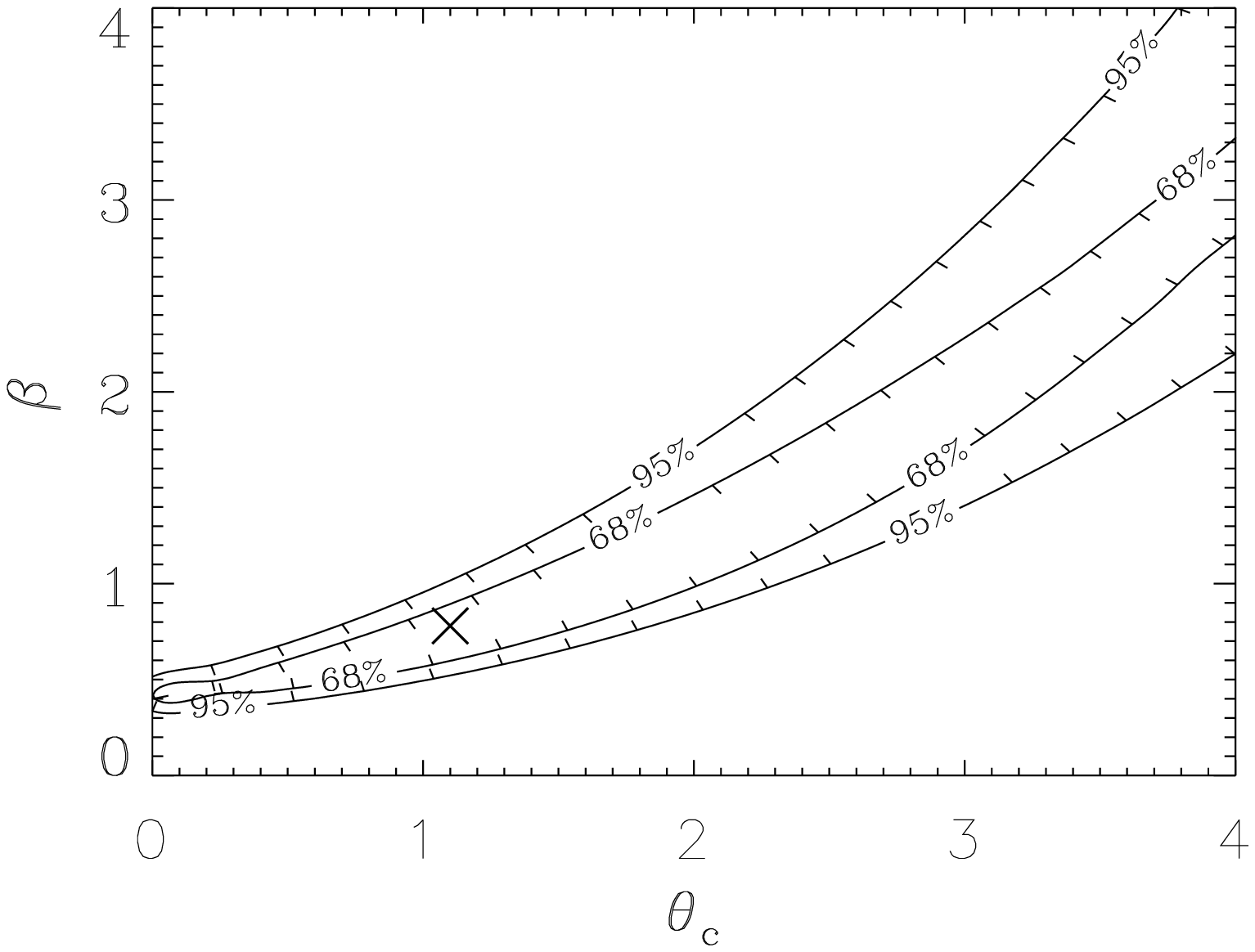}
\caption[]
{Isothermal density model parameter fits to the
coadded data scans across A1689.
Contours for $\theta_c$ and $\beta$ are shown at approximately
$68\%$ and $95\%$ confidence.
The ``X'' marks the best fit isothermal model parameter values
from the X-ray analysis.}
\label{rb1689}
\end{figure}

Fixing the position to the best fit value, we fit isothermal 
$\beta$ models for the 
surface brightness distribution to the coadded scans.
In Figure~\ref{rb1689},
We plot approximate confidence regions
for $\theta_c$ and $\beta$.
The observed S-Z surface brightness is consistent with
the best fit isothermal model parameters from the X-ray 
analysis and the assumption of an isothermal IC gas.

\subsection{$1.4\,$mm Observations}
\begin{figure}[htb]
\plotone{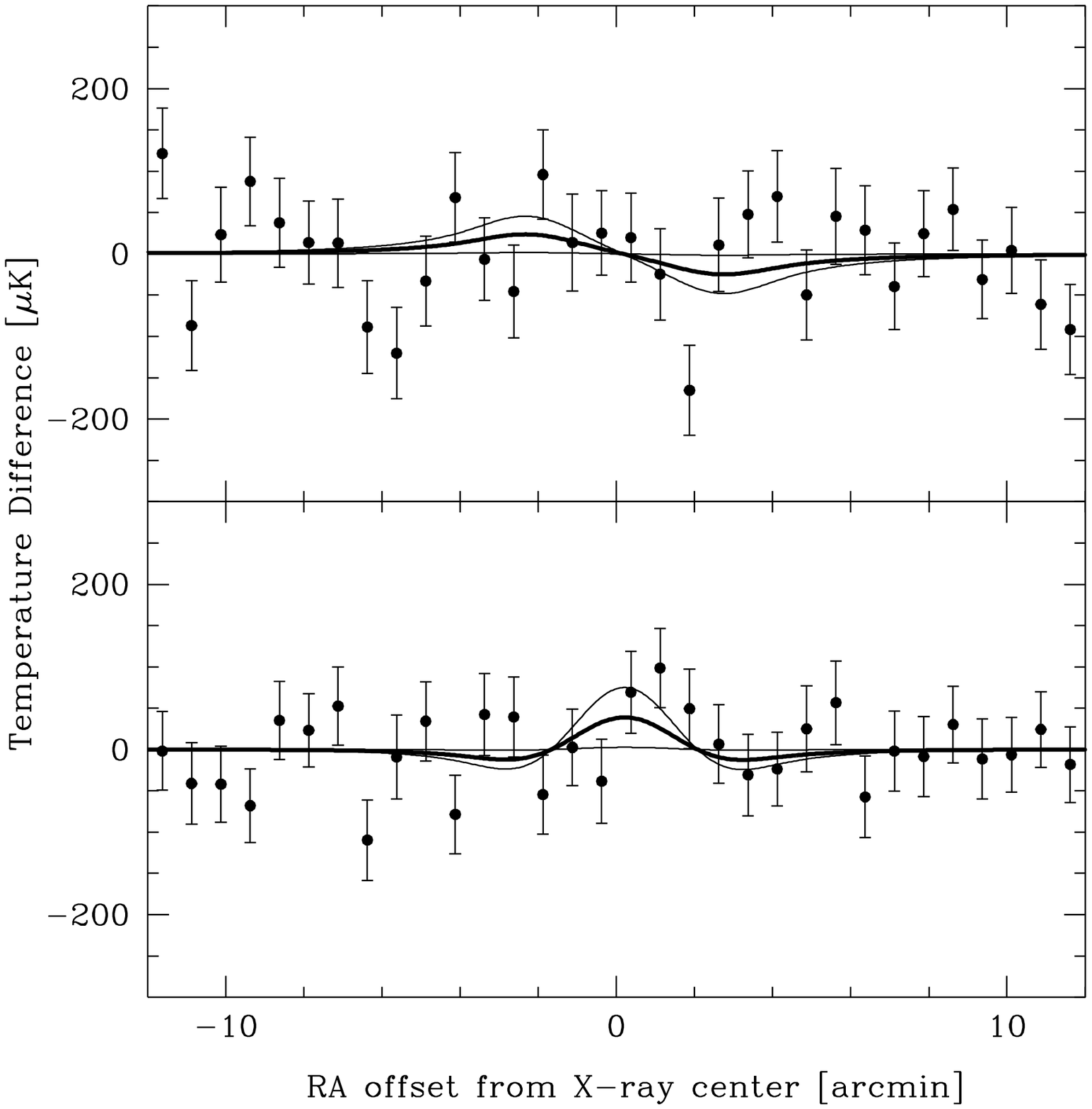}
\caption[]
{Coadded data for all April 1994 scans across A1689 at $1.4\,$mm.
The coaded scans from the two RAOs have been offset and
added to create a single coadded scan for each difference.
Both the $4.6^{\prime}$ difference (upper panel) and TBC
(lower panel) data are shown for the row
passing directly over the peak X-ray surface brightness.
The heavy line is the best fit isothermal model, and the lighter lines
represent the $1\sigma$ errors.
The scans are calibrated in terms of a Rayleigh-Jeans temperature
difference that completely fills one of the beams.}
\label{sb1689.14}
\end{figure}

We observed A1689 in the $1.4\,$mm band on the nights of
May 6-9 1994 for a total of $\sim 14\,$hours.
The scans were $30^{\prime}$ long with one row passing over
the X-ray center and the other $2.2\pr$ to the south.
The RAO of the X-ray center from the start of the scan was
alternated between $12^{\prime}$ and $18^{\prime}$
in sequential scans.
In Figure~\ref{sb1689.14}, we show the coadded $4.6\pr$ and
TBC scans over A1689.
Fitting the data to a model for the kinematic component of
the S-Z effect, we  
determine a marginally significant signal, the bulk of which
is due to the residual S-Z thermal component.

\subsection{Simultaneous Fits}
\begin{figure}[htb]
\plotone{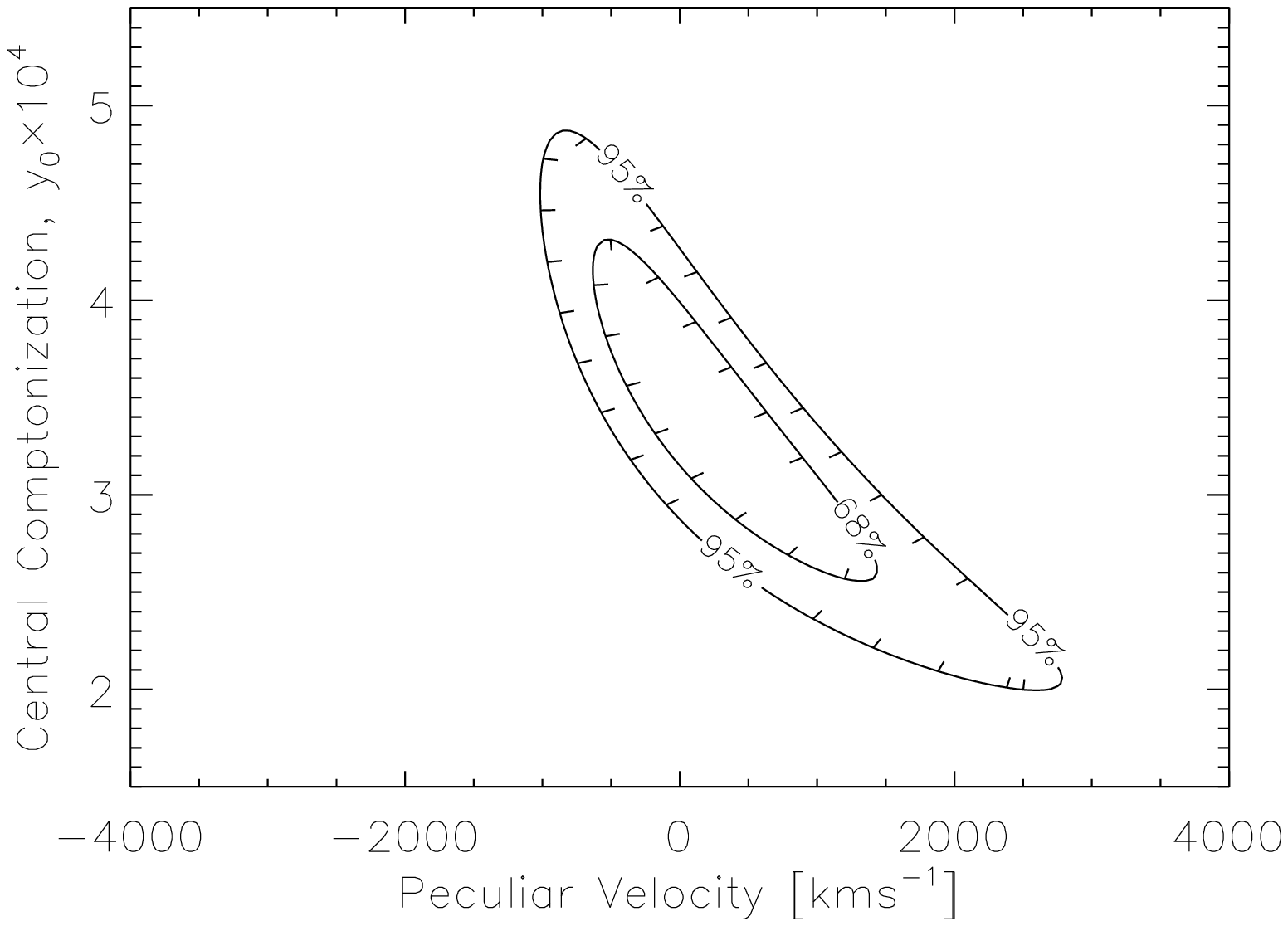}
\caption[]
{Results of isothermal ($T_e=8.2\,$keV) model fit to correlation
corrected corrected $2.1$ and $1.4\,$mm coadded scans over A1689.
The $68.3\%$ and $95.4\%$ confidence regions in
the peak \comp\ and radial component of the peculiar
velocity are shown.}
\label{yv1689}
\end{figure}

We simultaneously fit isothermal ($T_e=8.2\,$keV) models for the 
kinematic and thermal components of the S-Z effect to the
$2.1\,$ and $1.4\,$mm coadded scans.
The model positions are fixed to the best fit positions form the 
$2.1\,$mm fits; the model amplitudes, $y_0$ and $v_r$, are 
left as free parameters.
In Figure~\ref{yv1689},
we plot the $68.3\%$ and $95.4\%$ confidence
regions for $y_0$ and $v_r$.
The $68.3\%$ confidence intervals for the fit parameters 
considered individually are 
$y_0 = 3.43 \pm .59 \times 10^{-4}$ and
$v_{r} = +170^{+760}_{-570}\, {\rm kms}^{-1}$.
The results are unchanged for the allowed range of density 
model parameters.
We have repeated this analysis with the RAO differenced scans;
the results of the two analysis methods are compared
in Table~\ref{pvbt}.

We have determined $y_0$ and $v_r$ assuming two 
other values for the isothermal IC gas temperature which exceed the
range of temperatures allowed at $68.3\%$ confidence.
The peculiar velocities determined assuming $T_e=5\,$ and $10\,$keV 
are listed in Table~\ref{pvtt}.
The contribution of the uncertainty in the temperature
to the uncertainty in peculiar velocity 
($\D v_r < \pm 50\,{\rm kms}^{-1}$)
is much smaller than the statistical uncertainty in the measurement.

\section{Astrophysical Confusion}
\label{conf}
Although astrophysical confusion from randomly distributed sources is
expected to be small at mm wavelengths (\cite{FL}), the
possibility of confusion contributing a considerable systematic error
to the determination of $v_r$ in an individual cluster must be considered.
In the section, we examine the contribution of confusion,
including primary anisotropies of the CMB,
to the uncertainty in peculiar velocity for the two clusters considered.

\subsection{Radio Source Confusion}
\subsubsection{A2163}
\label{RC2163}
A VLA search towards A2163 shows evidence of a radio source $0.8\pr$ west
of the cluster center with an inverted spectrum (\cite{HB92}).
For this source, the flux rises from $1$ to $3\,$mJy between $6$ and
$2\,$cm suggesting a flux as large as $30\,$mJy at $2.1\,$mm.
Recently, this cluster has been imaged with the OVRO interferometer
operating at $30\,$GHz (\cite{Carlstrom}).
There is only one significant unresolved source found in the field.
It is centered at 
$16\ah,\,15\am,\,43.7\as;\; -06^{\circ},\,08\pr,\,45\2pr$ (J$2000$), 
close to the VLA source position, and has a flux of $1.4\,$mJy.
Due to the steeply falling spectra from $2\,$cm to $1\,$cm, the source flux is 
likely to be less than $1\,$mJy in any of the SuZIE bands.
However, this source is reported to be time variable (\cite{MBpc}).
Because of the elapsed time between measurements, the spectral index
and the inferred fluxes in the SuZIE bands are questionable.
The central region of the cluster has been imaged at $3.3\,$mm with
the IRAM interferometer.
Fischer \& Radford report an upper limit of $5\,$mJy
($2\sigma$ in a $20\2pr$ $\delta$, $10\2pr$ RA beam) on point
source emission within $1\pr$ of the X-ray center.
More sensitive observations with mm-wavelength interferometers are needed to
eliminate the possibility of confusion due to this particular radio source.
At present, we use the measured spectral index to infer that the flux in the
SuZIE bands is likely to be less than $1\,$mJy.

A2163 hosts the brightest radio halo yet discovered (\cite{HB95}).
From measurements at $1.5$ and $4.9\,$GHz, the integrated flux
from the radio halo is estimated to be less than $1\,$mJy in the
SuZIE $2.1\,$mm band (\cite{HPC}).
Taking the measured optical depth of A2163 and beam dilution into account,
a flux of $1\,$mJy in any of the 
three observing bands corresponds to an error of $\sim 120\, {\rm kms}^{-1}$ 
in the derived peculiar velocity for A2163.
The total uncertainty in the peculiar velocity of A2163 due to radio confusion 
is estimated to be $< \pm 170\,{\rm kms}^{-1}$.

\subsubsection{A1689}
\label{RC1689}
Radio emission from A1689 has been studied in detail by
Slee \ea (1994).
They find two bright sources near the cluster center which they label
\#13 and \#15a/b. 
Source \#13, at central position 
$13\ah,\,11\am,$ \\$30.2\as;$ $ -01^{\circ},\,20\pr,\,29\2pr$ (J$2000$), has a flux of 
$10.4\,$mJy and $2.75\,$mJy at frequencies of $1.5\,$GHz and $4.9\,$GHz.
Source \#15a/b, at central position 
$13\ah,\,11\am,\,31.4\as;$ $-01^{\circ},\,19\pr,\,33\2pr$ (J$2000$), has a flux of 
$41.1\,$mJy and $10.7\,$mJy at frequencies of $1.5\,$GHz and $4.9\,$GHz.
Using these results, they determine the spectral indices
to be $\alpha=-1.12$ and $\alpha=-1.14$ for sources 
\#13 and \#15a/b respectively.

Recently, sensitive maps of A1689 have been made with the OVRO 
interferometer operating at $30\,$GHz (\cite{Carlstrom}).
They find two significant point sources in the field with approximate
positions $13\ah,\,11\am,\,30.3\as;$ $-01^{\circ},\,20\pr,\,26\2pr$ (J$2000$) and
$13\ah,\,11\am,\,31.0\as;$ $-01^{\circ},\,19\pr,\,36\2pr$ (J$2000$) and 
fluxes $1.3$ and $0.4\,$mJy.  
Within the limits of the accuracy of the measurement ($\sim \pm 5\2pr$),
these positions are coincident with the previously identified
radio sources.
The fluxes are used to determine the spectral indices from
$4.9$ to $30.0\,$GHz.
We find $\alpha=-1.06$ and $\alpha=-1.14$ for sources \#13 and \#15a/b,
demonstrating that they continue their steep decrease in flux to $30\,$GHz.

Using the measured spectral indices, we estimate the flux in  
all three SuZIE bands
to be less than $0.3\,$mJy from the brighter of the two sources.
This is especially conservative in that the brighter of the two
sources is $\approx 1\pr$ north of the scan path, and therefore is never 
inside the FWHM of any of the array beams. 
The smaller beam dilution and higher optical depth of A1689
means that a $1\,$mJy point source in each of the three SuZIE bands produces 
only $\approx 63\,{\rm kms}^{-1}$ error.
We conclude that radio source confusion contributes $<\pm 19\,{\rm kms}^{-1}$
to the determination of the peculiar velocity in A1689.

\subsection{IR Cirrus}
We have used IRAS $100\,\mu$m sky maps to place limits on 
confusion due to IR cirrus emission associated with our galaxy. 
The images are obtained from the Infrared Sky Survey Atlas (ISSA) 
made available over the internet through the Infrared Processing 
and Analysis Center (IPAC). 
The images have had zodiacal light subtracted and
been ``de-striped''.
Each image is $2^{\circ}\times 2^{\circ}$ centered on the
peak X-ray Surface Brightness for each cluster.
They are composed of $1.5\pr$ square pixels,
although the actual resolution of IRAS at $100\,\mu$m is  
approximately $4\pr \times 5\pr$.

The line of sight to A2163 passes near the edge of the
galactic plane. 
In our scan region, the average $100\,\mu$m flux is 
$17.2\, {\rm MJy}\,{\rm sr}^{-1}$.  
The RMS variation of the flux in the $1.5\pr$ square 
pixels across the scan is $0.78\, {\rm MJy}\,{\rm sr}^{-1}$.
When the map is convolved with the SuZIE differential beam patterns, we 
find that the $100\,\mu$m differential flux at any point in the scan is 
$<\pm 126\, {\rm mJy}/$beam.
After the removal of a gradient in brightness.
this result holds for fits to both the $4.6\pr$ 
difference and the TBC.
In the analysis, we have not fixed the model position to a particular point 
in the scan, doing this will result in a lower limit. 

The line of sight to A1689 passes though a region of low
IR cirrus emission.
In our scan region, the average $100\,\mu$m flux is
$2.9\, {\rm MJy}\,{\rm sr}^{-1}$.  
The RMS variation of the flux in the image $1.5\pr$ square 
pixels is $0.16\, {\rm MJy}\,{\rm sr}^{-1}$.
When the map is convolved with the SuZIE beam patterns, we find that 
the $100\,\mu$m differential flux at any point in the scan is 
$<\pm 76\, {\rm mJy}/$beam.
This result appears to be dominated by residual striping in the map.
A more sophisticated analysis may result in significantly lower 
limits.

We scale the maps to our observing frequencies by assuming
the emission in the $100\,\mu$m map is due entirely to 
$23\,$K dust with an emissivity, $\epsilon \propto \nu^{-1.65}$ 
(\cite{Wright91}).
In Table~\ref{dust}, we list the upper limits on the differential 
flux from IR cirrus scaled in the SuZIE bands.
The flux limits are used to calculate the maximum confusion 
in peculiar velocity due to anisotropic dust.

\begin{table}[htb]
\begin{center}
\begin{tabular}{ccccc}
\multicolumn{5}{c}{Upper Limits for IR Cirrus Confusion}\\\tableline\tableline
 & & \multicolumn{3}{c}{Wavelength}\\
Cluster & & $2.1$ & $1.4$ & $1.1$ \\\tableline
$A2163$ & Flux $[{\rm mJy/beam}]$ & $0.17$ & $0.85$ & $1.50$\\
        & $\Delta v_{pec}\, [{\rm kms}^{-1}]$ & $20$ & $102$ & $180$\\
$A1689$ & Flux $[{\rm mJy/beam}]$ & $0.10$ & $0.51$ & $0.93$\\
        & $\Delta v_{pec}\, [{\rm kms}^{-1}]$ & $6$ & $32$ & $59$\\
\end{tabular}
\end{center}
\caption[]{Upper limits on peculiar velocity confusion due to
IR cirrus in the SuZIE bands.
A1689 was observed only in the $2.1$ and $1.4\,$mm bands.}
\label{dust}
\end{table}

Cluster member galaxies with slowly rising spectra could 
make a significant contribution to the uncertainty in peculiar velocity.
Observations with existing mm-wavelength interferometers could reduce this
uncertainty for a $\tau=.01$ cluster to $<100 {\rm kms}^{-1}$ in a 
single $1/2$ day observation.

\subsection{Primary Anisotropies}
It is also possible for the measurement of the S-Z effect to be confused
by the presence of primary CMB anisotropies.
The spectrum of these distortions is identical to that introduced by
the S-Z kinematic effect and, therefore, especially serious for the
determination of cluster peculiar velocities.
The confusion in $v_r$, due to primary anisotropies of amplitude $\D T/T$, is
$\D v_r = 100\, (.01/\tau)\,
[(\D T/T)/3 \times 10^{-6}]\, {\rm kms}^{-1}$.
Haehnelt and Tegmark (1996) have estimated the
confusion limits
from primary anisotropies to the determination of cluster peculiar velocities.
For $\Omega = 1$ (CDM) models with $\Omega_{baryon} = .01-.1$,
cluster optical depth $\tau=0.015$, and 
the SuZIE beam size, they find $|\D v_{pec}| < 300\,{\rm kms}^{-1}$.
This contribution to uncertainty as well as those due to foreground sources
are listed for both clusters in Table~\ref{evpec}.

\renewcommand{\arraystretch}{1.25}
\begin{table}[htb]
\begin{center}
\begin{tabular}{lrr}
\multicolumn{3}{c}{Peculiar Velocities and Uncertainties}\\\tableline\tableline
Uncertainty & A2163 & A1689\\\tableline
Statistical & $490^{+910}_{-730}$ & $170^{+760}_{-570}$\\
Calibration (Flux) & $^{+174}_{-174}$ & $^{+5}_{-5}$\\
Calibration (Spectral) & $^{+240}_{-240}$ & $^{+160}_{-160}$\\
Baseline & $^{+900}_{-0}$ & $^{+140}_{-0}$\\
IC Gas Temperature & $^{+20}_{-10}$ & $^{+20}_{-40}$\\
Thermal Structure & $^{+70}_{-0}$ & $-$\\
Radio Confusion & $^{+170}_{-170}$ & $^{+19}_{-19}$\\
IR Cirrus & $^{+180}_{-180}$ & $^{+32}_{-32}$\\
Primary Anisotropies & $^{+300}_{-300}$ & $^{+200}_{-200}$\\\tableline
Total & $490^{+1370}_{-880}$ & $170^{+815}_{-630}$\\
\end{tabular}
\end{center}
\caption[]{Peculiar velocities for A2163 and A1689 and all considered
contributions to the uncertainty.}
\label{evpec}
\end{table}
\renewcommand{\arraystretch}{1.00}

\section{Conclusion}
\label{conc}
The kinematic S-Z effect has the potential to be a singularly sensitive
probe of the peculiar velocities of distant galaxy clusters.
The realization of this potential requires the separation of
the thermal and kinematic components of the S-Z effect.
This can only be practically carried out by 
observations at mm and sub-mm wavelengths.
We have used observations of the S-Z effect in three
mm-wavelength bands to place limits
on the peculiar velocities of two distant clusters.

The cluster A2163 was observed in the $2.1$, $1.4$,
and $1.1\,$mm bands.
The $2.1\,$mm band observations are used to determine
the position and morphology of the peak S-Z surface brightness.
The positions of the X-ray and S-Z peak surface brightnesses
are consistent when uncertainties in the ROSAT/PSPC and SuZIE
pointing are included. 
The S-Z surface brightness is consistent with the X-ray derived density
model and the assumption of an isothermal IC gas.
From the $1.1\,$mm band observations, we report a
significant detection of the S-Z thermal effect
as an increment.
Simultaneously fitting the $2.1$, $1.4$, and $1.1\,$mm data
with the isothermal kinematic and thermal component
surface brightness models, we determine
$y_0 = 3.62 \pm .49 \times 10^{-4}$ and
$v_r = +490^{+910}_{-730}\, {\rm kms}^{-1}$
at $68.3\%$ confidence.
The uncertainties in $v_r$ due to the thermal structure 
of the IC gas are found to be much smaller than the statistical 
uncertainty.

The cluster A1689 was observed in the $2.1$ and
$1.4\,$mm bands.
The $2.1\,$mm band observations were used to determine
the position of the peak S-Z surface brightness.
The PSPC and S-Z peak surface brightnesses are nearly coincident.
Fits to the S-Z surface brightness indicate the observed
morphology is consistent with the density model determined
from the X-ray observations and isothermal IC gas.
Simultaneously fitting to the $2.1$ and $1.4\,$mm data,
we determine 
$y_0 = 3.43 \pm .59 \times 10^{-4}$ and
$v_{r} = +170^{+760}_{-570}\, {\rm kms}^{-1}$
at $68\%$ confidence.

Unlike the determination of $H_0$, the determination of cluster
peculiar velocities does not place
high demands on the quality of the X-ray data.
The results are constant over the allowed range of X-ray
derived density models.
For both clusters, we have determined $v_r$ over a broad range of X-ray temperature;
the value of the peculiar velocity
depends weakly on the assumed $T_e$.
Astrophysical confusion of the peculiar velocity measurements
in the two clusters is likely to be dominated by primary 
anisotropies of the CMB at a level $\lesssim 300\,{\rm kms}^{-1}$. 
All the sources of uncertainty to the determination of
cluster peculiar velocities, that we have considered,
are listed in Table~\ref{evpec}. 
Including uncertainties due to the uncertainty in the 
IC gas temperature, thermal structure, density model, calibration,
and astrophysical confusion, we find
$v_r=+490^{+1370}_{-880}\,{\rm kms}^{-1}$ for A2163, and
$v_r=+170^{+815}_{-630}\,{\rm kms}^{-1}$ for A1689,
both at $68\%$ confidence.

Even though we have not yet reached sufficient sensitivity to
determine individual peculiar velocities, it is possible
to use results like those presented here, to probe the velocity
field on very large scales.
For example, Lauer \& Postman (1994) (LP) report a bulk
flow of galaxy clusters within a sphere of $15,000\,{\rm kms}^{-1}$ of
$v_{LP} = +730 \pm 174\, {\rm kms}^{-1}$.
A2163 and A1689 lie along lines of sight separated
from the direction of the LP bulk flow by only $\approx 29^{\circ}$ and
$\approx 17^{\circ}$ respectively.
If we assume that all the radial peculiar velocity of the clusters is
due to a bulk flow in the direction of the LP results, we find 
$v_{bulk} \approx  +280^{+750}_{-550}\, {\rm kms}^{-1}$ at $68.3\%$
confidence.
The results of this work cannot rule out
the continuation of the LP bulk flow to $z\sim .2$.
A modest increase in sensitivity coupled with the observation of clusters 
in the opposite direction to the LP flow (as a systematic check) 
would make a definitive test possible.

The sensitivity of the S-Z observations described here are limited 
by differential atmospheric emission.
It should be possible to use the distinct spectral signatures of
the S-Z effects to separate them from atmospheric noise.
This requires simultaneous measurements along a given line of sight
in at least three mm-wavelength bands.
We have constructed and begun observations with
an instrument capable of such measurements.
This instrument should eventually have the capability
of determining cluster peculiar velocities limited by the 
astrophysical confusion of the S-Z effect.

We would like to thank Ken Ganga for many informative discussions.
The presentation has also benefited significantly from the comments of an 
anonymous referee. 
Thanks to Antony Schinkel and the entire staff of the CSO for their
excellent support during the observations.
The CSO is operated by the California Institute of Technology under
funding from the National Science Foundation, contract \#AST-93-13929.
This work has been made possible by a grant from the David and Lucile
Packard foundation and by a National Science Foundation grant
\#AST-95-03226.

\markright{REFERENCES}

\end{document}